\documentclass[12pt]{article}
\pdfoutput=1
\usepackage{jheppub}
\usepackage{amsmath}
\usepackage{amsfonts}
\usepackage{amssymb}
\usepackage{graphicx}

\usepackage[export]{adjustbox}
\newcommand{\bmat}{\left(\begin{array}}
\newcommand{\emat}{\end{array}\right)}

\def\yzero{\smash{\hbox{$y\kern-4pt\raise1pt\hbox{${}^\circ$}$}}}

\def\beq{\begin{equation}}
\def\eeq{\end{equation}}
\def\beqa{\begin{eqnarray}}
\def\eeqa{\end{eqnarray}}

\def\-{\hphantom{-}}
\def\ov{\overline}
\def\s2{\frac{1}{\sqrt2}}

\def\beq{\begin{equation}}
\def\eeq{\end{equation}}
\def\beqa{\begin{eqnarray}}
\def\eeqa{\end{eqnarray}}

\def\IF{\relax{\rm I\kern-.18em F}}
\def\II{\relax{\rm I\kern-.18em I}}

\def\Dsl{\,\raise.15ex\hbox{/}\mkern-13.5mu D} 

\def\IC{{\bf C}}
\def\IS{{\bf S}}
\def\IR{{\bf R}}
\def\IZ{{\bf Z}}
\def\IX{{\bf X}}
\def\IY{{\bf Y}}

\def\IT{{\bf T}}
\def\IP{\bf P}
\def\IRP{{\bf RP}}

\def\NN{{\cal N}}

\def\C{{\bf C}}

\newcommand{\drawsquare}[2]{\hbox{%
\rule{#2pt}{#1pt}\hskip-#2pt
\rule{#1pt}{#2pt}\hskip-#1pt
\rule[#1pt]{#1pt}{#2pt}}\rule[#1pt]{#2pt}{#2pt}\hskip-#2pt
\rule{#2pt}{#1pt}}

\newcommand{\fund}{~\raisebox{-.5pt}{\drawsquare{6.5}{0.4}}~}
\newcommand{\antifund}{~\overline{\raisebox{-.5pt}{\drawsquare{6.5}{0.4}}}~}

\newcommand{\symm}{~\raisebox{-.5pt}{\drawsquare{6.5}{0.4}}\hskip-0.4pt%
        \raisebox{-.5pt}{\drawsquare{6.5}{0.4}}~}

\newcommand{\asymm}{~\raisebox{-3.5pt}{\drawsquare{6.5}{0.4}}\hskip-6.9pt%
        \raisebox{3pt}{\drawsquare{6.5}{0.4}}~}

\newcommand{\antisymm}{~\overline{\raisebox{-.5pt}{\drawsquare{6.5}{0.4}}\hskip-0.4pt%
        \raisebox{-.5pt}{\drawsquare{6.5}{0.4}}}~}

\catcode`\@=11   
\newdimen\@rotdimen
\newbox\@rotbox  

\def\@vspec#1{\special{ps:#1}}
\def\@rotstart#1{\@vspec{gsave currentpoint currentpoint translate
   #1 neg exch neg exch translate}}
\def\@rotfinish{\@vspec{currentpoint grestore moveto}}
\def\@rotr#1{\@rotdimen=\ht#1\advance\@rotdimen by\dp#1%
   \hbox to\@rotdimen{\hskip\ht#1\vbox to\wd#1{\@rotstart{90 rotate}%
   \box#1\vss}\hss}\@rotfinish}
\def\@rotl#1{\@rotdimen=\ht#1\advance\@rotdimen by\dp#1%
   \hbox to\@rotdimen{\vbox to\wd#1{\vskip\wd#1\@rotstart{270 rotate}%
   \box#1\vss}\hss}\@rotfinish}%
\def\@rotu#1{\@rotdimen=\ht#1\advance\@rotdimen by\dp#1%
   \hbox to\wd#1{\hskip\wd#1\vbox to\@rotdimen{\vskip\@rotdimen
   \@rotstart{-1 dup scale}\box#1\vss}\hss}\@rotfinish}%
\def\@rotf#1{\hbox to\wd#1{\hskip\wd#1\@rotstart{-1 1 scale}%
   \box#1\hss}\@rotfinish}%
\def\rotate{\@ifnextchar[{\@rotate}{\@rotate[l]}}
\def\@rotate[#1]#2{\setbox\@rotbox=\hbox{#2}\@nameuse{@rot#1}\@rotbox}

\catcode`\@=12

\begin{document}

\makeatletter
\@addtoreset{equation}{section}
\makeatother
\renewcommand{\theequation}{\thesection.\arabic{equation}}
\pagestyle{empty}
\rightline{ IFT-UAM/CSIC-18-102}
\vspace{1.2cm}
\begin{center}
\LARGE{\bf Supersymmetry Breaking Warped Throats and the Weak Gravity Conjecture \\[12mm] }
\large{Ginevra Buratti$^{1}$, Eduardo Garc\'{\i}a-Valdecasas$^{1,2}$,  Angel M. Uranga$^1$\\[4mm]}
\footnotesize{${}^{1}$ Instituto de F\'{\i}sica Te\'orica IFT-UAM/CSIC,\\[-0.3em] 
C/ Nicol\'as Cabrera 13-15, 
Campus de Cantoblanco, 28049 Madrid, Spain} \\ 
${}^2$ Departamento de F\'{\i}sica Te\'orica, Facultad de Ciencias\\[-0.3em] 
Universidad Aut\'onoma de Madrid, 28049 Madrid, Spain\\

\vspace*{5mm}

\small{\bf Abstract} \\
\end{center}
\begin{center}
\begin{minipage}[h]{17.0cm}
We generalize the swampland criterion forbidding stable non-supersymmetric AdS vacua and propose a new swampland conjecture forbidding stable non-supersymmetric ``locally AdS'' warped throats. The conjecture is motivated by the properties of systems of fractional D3-branes at singularities, and can be used to rule out large classes of warped throats with supersymmetry breaking ingredients, and their possible application to de Sitter uplift. In particular, this allows to reinterpret the runaway instabilities of the gravity dual of fractional branes in the dP$_1$ theory,  and to rule out warped throats with Dynamical Supersymmetry Breaking D-brane sectors at their bottom. We also discuss the instabilities of warped throats with supersymmetry broken by the introduction of anti-orientifold planes. These examples lead to novel decay mechanisms in explicit non-supersymmetric examples of locally AdS warped throats, and also of pure AdS backgrounds.
\end{minipage}
\end{center}
\newpage
\setcounter{page}{1}
\pagestyle{plain}
\renewcommand{\thefootnote}{\arabic{footnote}}
\setcounter{footnote}{0}

\tableofcontents

\vspace*{1cm}
\section{Introduction: Quantum Gravitational String Phenomenology}

The recent flurry of activity, largely triggered by \cite{delaFuente:2014aca,Rudelius:2015xta,Montero:2015ofa},  in constraining phenomenological string model building using Quantum Gravity swampland criteria \cite{Vafa:2005ui,Ooguri:2006in,ArkaniHamed:2006dz,Ooguri:2016pdq,Freivogel:2016qwc,Obied:2018sgi,Harlow:2018jwu,Harlow:2018tng} (see \cite{Brennan:2017rbf} for a recent review) is giving birth to an emerging field, which can deservedly claim the designation of \textbf{Quantum Gravitational String Phenomenology}.

The application of constraints convincingly argued to hold in any theory of Quantum Gravity is leading to new breakthroughs. In particular, the Weak Gravity Conjecture (WGC) \cite{ArkaniHamed:2006dz} (see \cite{Cheung:2014vva,delaFuente:2014aca,Rudelius:2015xta,Montero:2015ofa,Brown:2015iha,Brown:2015lia,Heidenreich:2015wga,Hebecker:2015rya,Bachlechner:2015qja,Junghans:2015hba,Ibanez:2015fcv,Hebecker:2015zss,Heidenreich:2016aqi,Montero:2016tif,Ibanez:2017oqr,Gonzalo:2018tpb,Gonzalo:2018dxi} for different formulations and applications) has motivated the remarkable statement that stable non-supersymmetric  Anti de Sitter (AdS) vacua are not possible in Quantum Gravity \cite{Ooguri:2016pdq,Freivogel:2016qwc}. This AdS-WGC constraint is largely motivated by the application of the refined WGC to systems of branes in the near horizon limit, and has received direct support from the study of decays of non-supersymmetric AdS vacua in string theory via bubbles of nothing \cite{Ooguri:2017njy}. The AdS-WGC has been argued to have far-reaching implications for particle physics and its scales \cite{Ibanez:2017oqr,Gonzalo:2018tpb,Gonzalo:2018dxi}.

There are also recent proposals of swampland criteria attempting to rule out de Sitter vacua as well \cite{Obied:2018sgi,Garg:2018reu,Ooguri:2018wrx}, possibly in certain regimes under parametric control. This claim clashes with familiar roadmaps for the construction of de Sitter vacua in string theory \cite{Kachru:2003aw,Balasubramanian:2005zx}, see \cite{Cicoli:2018kdo,Kachru:2018aqn} for recent discussion. A key ingredient in the parametric control of these scenarios is the presence of warped throats \cite{Klebanov:2000hb,Giddings:2001yu} at whose bottom the supersymmetry breaking sectors are localized, so that they undergo a redshift crucial for the tunability of the 4d vacuum energy. Starting from the original proposal of supersymmetry breaking by anti-D3-branes \cite{Kachru:2003aw}, there is a rich variety of proposals, see e.g. \cite{Burgess:2003ic,Kallosh:2015nia,Retolaza:2015nvh}. Hence, it is interesting to explore the interplay of non-supersymmetric warped throats with constraints from Quantum Gravity.

In this paper we consider non-compact warped throats and constrain these 5d backgrounds by proposing a new swampland conjecture, the {\em local AdS-WGC}, which generalizes the AdS-WGC to locally AdS warped throats. The conjecture is motivated by considering the near horizon limit of systems of fractional D-branes at singularities, but should hold more in general. Although it does not constrain metastable non-supersymmetric throats, hence has no direct implication for e.g. anti-D3-brane models, it can be used to rule out large classes of warped throats with supersymmetry breaking sectors at their bottom. We study this phenomenon in several explicit examples, shedding new light on already known instabilities in supersymmetry breaking D-brane models, such as the dP$_1$ theory, and unveiling novel decay channels in AdS or locally AdS backgrounds. For instance, we explicitly discuss warped throats with supersymmetry broken by the introduction of anti-orientifold planes.

A remarkable feature of these examples is that the non-supersymmetric backgrounds are stable at the classical level, and that the pathologies arise at the quantum level, often by nucleation of bubbles hosting interiors of more stable vacua. This is consistent with the interpretation of these constraints as arising from consistency in Quantum Gravity.

The paper is organized as follows: In Section \ref{sec:dimer-intro} we review systems of D-branes at singularities and fractional branes using the powerful toolkit of dimer diagrams. In Section \ref{sec:local-ads-wgc} we propose the local AdS-WGC criterion; we derive it in section \ref{sec:derivation-deformed}, and use it in section \ref{sec:deformation-dsb} to reinterpret the properties of supersymmetric and non-supersymmetric warped throats dual to fractional D3-branes in toric singularities. In section \ref{sec:metastable} we discuss the situation for throats with meta-stable supersymmetry breaking. In Section \ref{sec:dsb-throat} we consider an illustrative example of a system of D3-branes with Dynamical Supersymmetry Breaking due to strong dynamics and consider its embedding into warped throats. The D-brane gauge theory is discussed in section \ref{sec:dsb-gauge}, and in section \ref{sec:dsb-ads} we describe the instabilities that arise when embedded into AdS or locally AdS warped throats, in agreement with the (local) AdS-WGC implications for non-supersymmetric throats; in section \ref{sec:ntwo} we describe the local AdS-WGC statement in an explicit example illustrating how it applies to non-supersymmetric throats from $\NN=2$ fractional branes. Section \ref{sec:anti-oplanes} treats warped throats with supersymmetry broken by the presence of anti-orientifold-planes. In section \ref{sec:anti-oplanes-throats} we discuss generalities about such throats. In section \ref{sec:anti-oplanes-o3} we focus on anti-O3-planes, describe their different kinds and their interaction with systems of D3-branes. In section \ref{sec:anti-oplanes-o3-throats} we discuss the corresponding gravitational backgrounds and describe their instabilities, in agreement with the (local) AdS-WGC statement.
Finally, in Section \ref{sec:conclu} we give our conclusions.

\section{Review of dimers and fractional branes}
\label{sec:dimer-intro}

Here we briefly review some ingredients of the dimer diagram description of D3-branes at singularities. The initiated reader is welcome to skip it and jump into the next sections.

\begin{figure}[htb]
\begin{center}
\includegraphics[scale=.4]{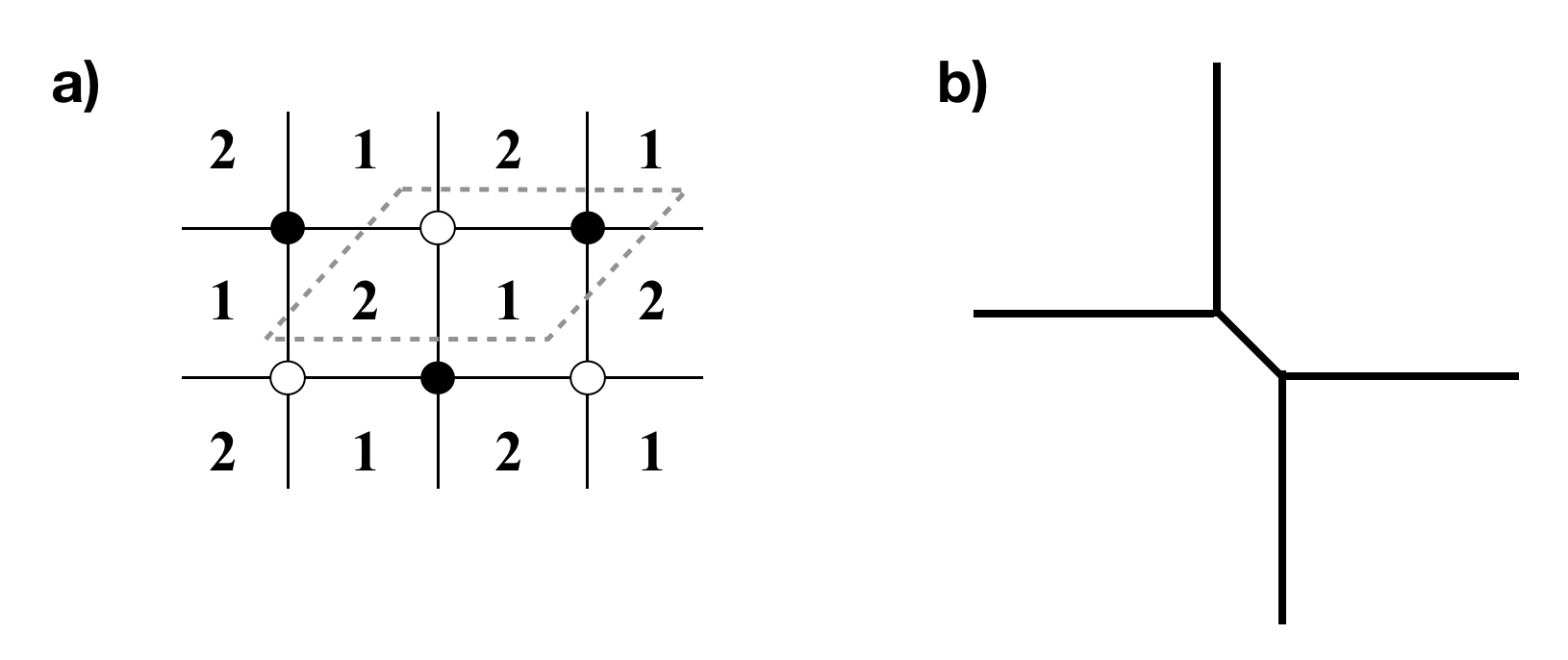} 
\caption{\small (a) Dimer diagram for the theory of D-branes at a conifold. The dashed line is the unit cell in the periodic array. (b)  Web diagram of the conifold. We have displayed it with a finite size $\IS^2$ (middle segment) for clarity; the actual singularity arises when this $\IS^2$ is blown-down.}
\label{fig:conifold-dimer2}
\end{center}
\end{figure}

The gauge theories on D3-branes at toric CY threefold singularities are nicely encoded in a combinatorial graph known as dimer diagram \cite{Franco:2005rj,Franco:2005sm} (see also \cite{Kennaway:2007tq,Yamazaki:2008bt} and references therein). They are (bipartite) graph tilings of $\IT^2$, or equivalently infinite periodic graphs in $\IR^2$. Their faces correspond to gauge factors, edges represent  bi-fundamental chiral multiplets (oriented e.g. clockwise around black nodes, and counterclockwise around white nodes), and nodes represent superpotential couplings (with sign determined by the node color). As an illustration, the diagram for the conifold is shown in figure \ref{fig:conifold-dimer2}(a).
The corresponding gauge theory \cite{Klebanov:1998hh} has gauge group $U(n_1)\times U(n_2)$, bi-fundamental chiral multiplets in two copies of the representation $(\fund_1,\antifund_2)+(\antifund_1,\fund_2)$, denoted by $A_i$, $B_i$, $i=1,2$, and a superpotential $W=\epsilon_{ik}\epsilon_{jl} A_iB_jA_kB_l$.
 
The geometric information about the CY singularity is encoded in simple combinatorial objects in the dimer, whose discussion we skip, directing the interested reader to the references. We just mention that the geometries are encoded in web diagrams, which specify the fibration structure of the corresponding toric geometry. The web diagram can be obtained by constructing the zig-zag paths in the dimer (these are paths constructed out of sequences of edges which turn maximally left at black nodes and maximally right at white nodes) and translating the non-trivial $(p,q)$ windings of the path on the two non-trivial 1-cycles in $\IT^2$ into the $(p,q)$ labels of external legs in the web diagram. The web diagram for the conifold is shown in figure \ref{fig:conifold-dimer2}(b).
 %

The choice of ranks $n_i$ in the gauge groups of the dimer theories is arbitrary, but constrained by cancellation of RR tadpoles. These are equivalent to cancellation of non-abelian gauge anomalies (understood as formally imposed for all gauge factors, even those of possible empty faces). These conditions also guarantee the cancellation of mixed $U(1)$ anomalies thanks to Green-Schwarz couplings. There are in fact topological $BF$ couplings with RR 2-forms making all $U(1)$ factors massive (even the non-anomalous ones, see \cite{Ibanez:1998qp}). Supersymmetry of the configuration implies that blow-up modes couple as (field dependent) FI terms to the D3-branes. Although these $U(1)$'s are massive, it still makes sense to discuss them if the corresponding couplings to localized closed string modes are taken into account.

The choice of all ranks equal $n_i\equiv N$ for all $i$ is always allowed, and corresponds to D3-branes which can move off the singularity, as signaled by corresponding flat directions in the D3-brane gauge theory. This in fact underlies the way in which the dimer encodes the CY threefold geometry, as the moduli space of  a single such D-brane. These D3-branes are referred to as dynamical, or regular (since, for orbifold singularities, they are associated to the regular representation of the orbifold group \cite{Douglas:1996sw}).

\begin{figure}[!htp]
\begin{center}
\includegraphics[scale=0.3]{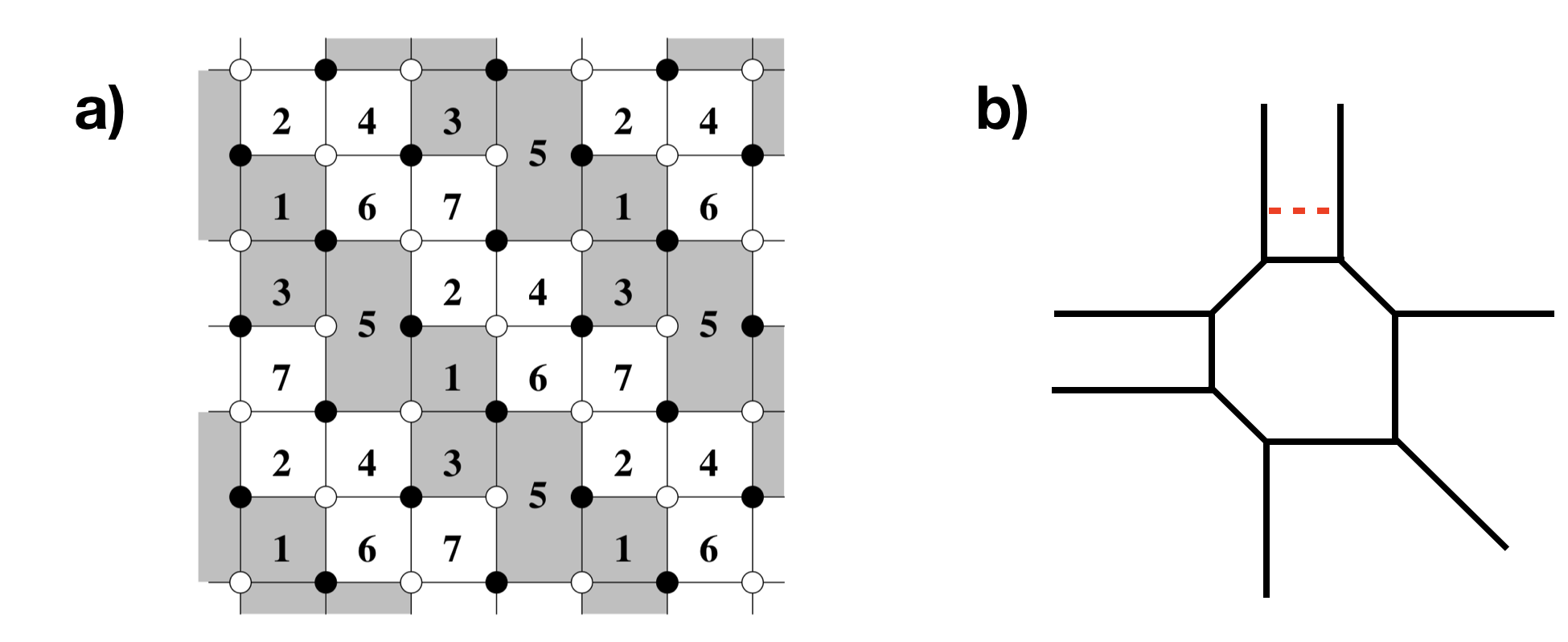}
\caption{a) Dimer diagram showing an $\NN=2$ fractional brane in the PdP$_4$ theory; b) Web diagram, displaying the corresponding mobile $\IP_1$ as a red discontinuous segment.}
\label{Nequalstwo}
\end{center}
\end{figure}

Other rank assignments consistent with the RR tadpole constraints are known as fractional branes. They can be regarded as D5-branes wrapped on 2-cycles (collapsed at the singularity) such that their dual 4-cycle is non-compact. This allows the RR charge carried by the D-branes to escape to infinity. These can always be written as combinations of certain basis of fractional branes, which fall into different classes, as described in \cite{Franco:2005zu}, as follows: 

\smallskip

$\bullet$ The so-called $\NN=2$ fractional branes correspond to an overall increase of ranks in a subset of faces bounded by zig-zag paths associated to the same $(p,q)$ 1-cycle in the dimer $\IT^2$. They are associated to parallel external legs in the web diagram, or equivalently to curves of $\IC^2/\IZ_k$ singularities sticking out of the singularity at the origin. The gauge theory on these fractional D3-branes has a flat direction, parametrized by the meson obtained by concatenation of bifundamentals joining the faces bounded by the zig-zag paths in the dimer. The flat direction describes the possibility of moving the fractional D-brane off the origin along the curve of singularities, to become a fractional brane of $\IC^2/\IZ_k$, namely a D5-brane wrapped on one of the collapsed 2-cycles of the orbifold singularity. The gauge theory on this branch is the $\NN=2$ $A_{k-1}$ quiver gauge theory \cite{Douglas:1996sw}, hence the name. An example of $\NN=2$ fractional brane is shown in Figure \ref{Nequalstwo}.

\smallskip

\begin{figure}[!htp]
\begin{center}
\includegraphics[scale=0.4]{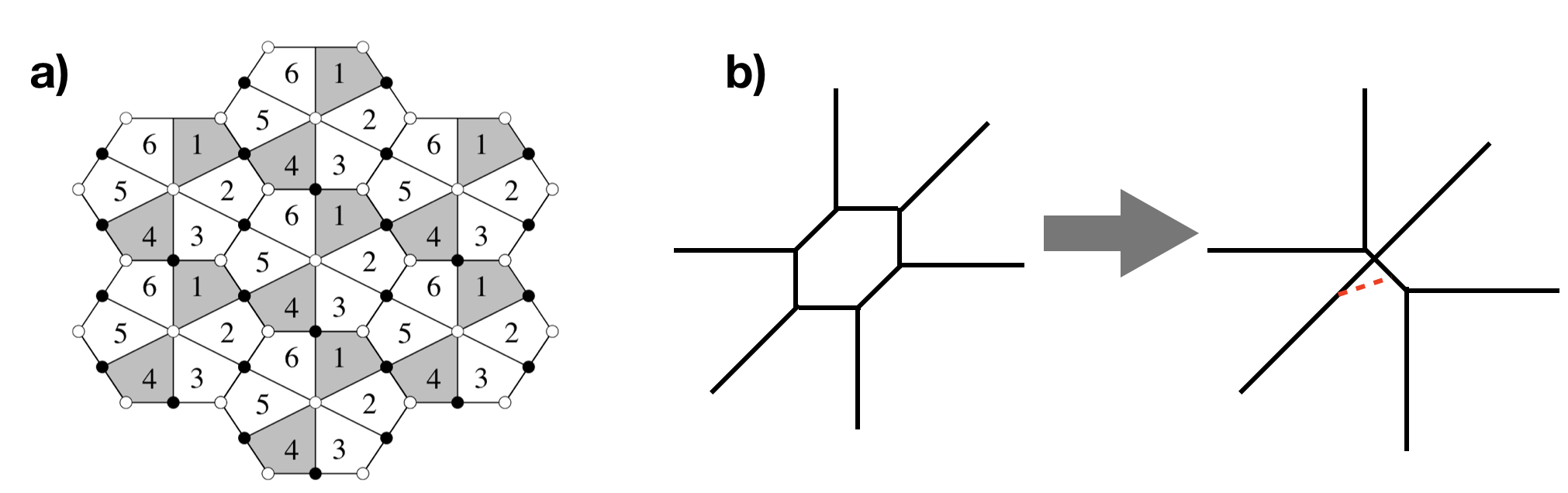}
\caption{a) Dimer diagram showing a deformation fractional brane in the dP$_3$ theory; b) Web diagram, and its splitting into subwebs in equilibrium, with the finite size $\IS^3$ displayed as a red discontinuous segment suspended between the subwebs.}
\label{deformation-brane}
\end{center}
\end{figure}

$\bullet$ The so-called deformation branes are associated to complex deformations of the CY threefold singularity. They are associated to splittings of the web diagram into sub-webs in equilibrium. The rank assignment corresponds to an overall increase of ranks in the subset of faces bounded by the splitting. Namely, the homological sum of the zig-zag paths associated to the sub-web removed (in a given complex deformation, the two sub-webs give the same result, due to the condition that the total sum of $(p,q)$ charges for external legs is zero). They correspond to checkerboard pictures on the dimer. The complex deformation of the geometry has a field theory counterpart, in which the gauge theory on the fractional branes confines and has a complex deformed moduli space. The resulting gauge theories are associated to the two sub-webs \cite{Franco:2005fd,GarciaEtxebarria:2006aq}. An example of a deformation fractional brane is shown in Figure \ref{deformation-brane}.

The gauge theory arising from a set of $N$ regular D3-branes and $M$ (deformation) fractional branes, leads to RG flows with a sequence of Seiberg duality cascades, along which the overall number of D3-branes $N$ is reduced in multiples of $M$, and the number $M$ of D5-branes remains fixed. The gravity dual corresponds to a warped throat supported by 
RR 3-form fluxes on the 3-cycles associated to the complex deformed singularity, and NSNS flux in the dual (non-compact) 3-cycle. Their combination $G_3=F_3-\tau H_3$ is an ISD 3-form of type $(2,1)$, thus preserving supersymmetry \cite{Becker:1996gj,Dasgupta:1999ss,Grana:2000jj}. The throat is locally similar to AdS$_5\times \IX_5$, but with logarithmic changes in the cosmological constant and the RR 5-form flux along the radial direction.

The simplest example is the conifold, studied exhaustively in \cite{Klebanov:2000hb} both from the viewpoints of field theory and of its gravity dual warped throat. The generalization of duality cascades in gauge theories associated with fractional branes in more general singularities has been studied in \cite{Franco:2004jz,Franco:2005fd}. We will consider the gravity dual of deformation branes in general singularities in section \ref{sec:derivation-deformed}.

$\bullet$  The last class corresponds to the remaining kind of fractional branes. Their corresponding rank assignments on faces have no correspondence with a set of zig-zag paths defining a sub-web in equilibrium. Therefore, there is no geometric complex deformation of the singularity associated to them. Indeed, contrary to deformation fractional branes, their infrared dynamics involves non-abelian gauge dynamics (even for the minimal such fractional brane) and results in the absence of a supersymmetric vacuum (hence they were dubbed DSB branes in \cite{Franco:2005zu}, see also \cite{Berenstein:2005xa,Bertolini:2005di}). On the other hand, similarly to deformation fractional branes, they can trigger duality cascades in the presence of $N$ regular D3-branes, which define some warped throats (albeit with naked singularities in the infrared region) \cite{Franco:2004jz}. The discussion of the infrared dynamics, supersymmetry breaking, and its implications for the gravity dual and the deformed AdS-WGC are discussed in section \ref{sec:derivation-deformed}

\medskip

\begin{figure}[htb]
\begin{center}
\includegraphics[scale=.4]{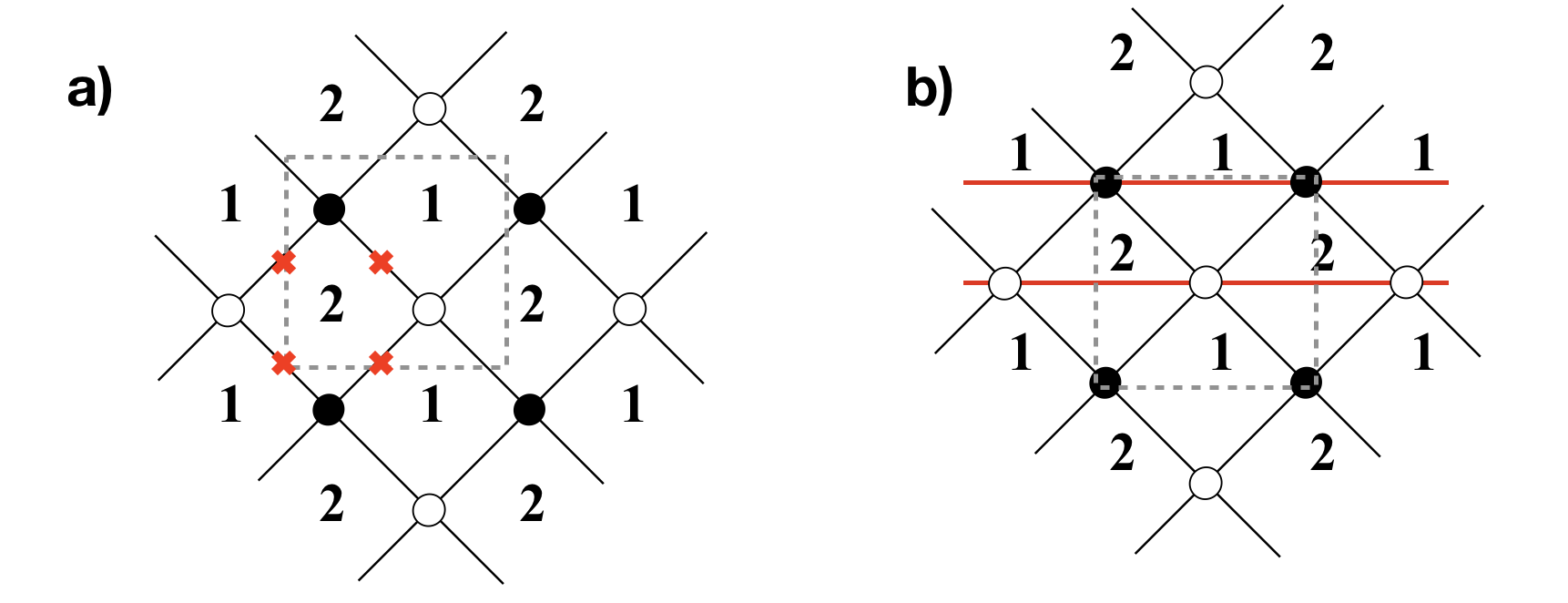} 
\caption{\small Dimer diagrams for orientifolds of the conifold with fixed points (a) or fixed lines (b).}
\label{fig:coni-orientifold}
\end{center}
\end{figure}

In this paper we will also exploit systems of D-branes at orientifolds of toric singularities. They can be usefully encoded in suitable modifications of dimer diagrams. The general description was provided in \cite{Franco:2007ii}, and corresponds to modding out the dimer diagram by a $\IZ_2$ involution. There are two kinds of orientifold quotients, classified by their fixed sets being lines or points. Two such orientifolds of the conifold theory are shown in figure \ref{fig:coni-orientifold}. It is easy to construct other examples, see later and \cite{Franco:2007ii}.

In the following we will mainly focus on models with orientifold fixed points in the dimer. For this class, the rules are as follows (see \cite{Franco:2007ii} for detailed derivations). Each orientifold point carries a $\pm$ sign, with the constraint that the number of orientifold planes with the same sign is even (resp. odd) for dimers with number of nodes given by $4k$ (resp. $4k+2$). Orientifold points with charge $+$ (resp $-$) in the middle of a dimer face project down the corresponding gauge factor to $SO(n_a)$ (resp. $USp(n_a)$ ). Orientifold points with charge $+$ (resp. $-$) in the middle of a dimer edge project down the corresponding bifundamental onto the two-index symmetic (resp. antisymmetric) representation. Finally, faces and edges not mapped to themselves by the orientifold, combine with their images and descend to $U(n_a)$ gauge factors and bi-fundamental matter multiplets in the orientifold theory. 

\section{The local AdS-WGC swampland criterion}
\label{sec:local-ads-wgc}

\subsection{Derivation}
\label{sec:derivation-deformed}

The WGC \cite{ArkaniHamed:2006dz}, in its minimal formulation establishes that in any theory including quantum gravity, any $U(1)$ gauge factor should have a super-extremal charged particle, namely $q\geq m$, in natural units. This has been generalized to other $p$-form gauge fields, requiring the existence of the corresponding branes with tensions bounded by their charges, $Q\geq T$, an extension natural in string theory models via T-duality. 

The proposal in \cite{Ooguri:2016pdq} of a {\em refined} WGC establishes that the inequality is saturated only for BPS states in supersymmetric theories. This further motivates the {\em AdS-WGC} statement that theories of quantum gravity do not have stable non-supersymmetric AdS vacua, which are thus in the swampland, rather than the string landscape. The AdS-WGC is largely motivated by a particular (but large) class of AdS backgrounds in string theory, which correspond to flux compactifications arising as near horizon limits of systems of D-branes. A prototypical example is the type IIB AdS$_5\times \IS^5$ solution with $N$ units of RR 5-form flux on the $\IS^5$, which arises as the near horizon limit of a system of $N$ D3-branes in flat 10d spacetime \cite{Maldacena:1997re}.
In short, the $T=Q$ condition is crucial in the structure of these vacua, in which the tension creating the spacetime curvature is balanced against the flux sourced by the brane charge in the underlying picture. This proposal is further supported by the study of instabilities of non-supersymmetric AdS vacua due to bubbles of nothing \cite{Ooguri:2017njy}. The AdS-WGC  is a powerful statement, which e.g. has subsequently been applied to derive novel constraints on particle physics \cite{Ibanez:2017kvh,Ibanez:2017oqr,Gonzalo:2018tpb,Gonzalo:2018dxi}.

In this paper we propose a generalization of the conjecture, which we dub the {\em local AdS-WGC}. It states that certain warped throats backgrounds, which are  AdS locally in the radial direction but have a slow variation of the local 5d value of the cosmological constant, are not consistent in quantum gravity, except for supersymmetric cases. The precise formulation will be manifest from the derivation below.

The derivation follows the strategy of \cite{Ooguri:2016pdq} for AdS fluxed backgrounds, by taking a near horizon limit of D-brane systems. In our case, we apply the near horizon description to systems of regular and fractional D3-branes at singularities, in particular the toric CY singularities of section \ref{sec:dimer-intro}. We note that the discussion below also applies to throats from $\NN=2$ fractional branes, despite the presence of singularities in the near horizon geometry, if one accounts for the additional fields from the twisted sectors, see \ref{sec:ntwo} for extra details. 

The backgrounds correspond to the holographic duals of (the UV regime of) gauge theories with cascading RG flows, like the familiar conifold example. The statements below have well-established translations to the holographic dual gauge theory on the D-branes, but we prefer to emphasize the properties of the gravity side.

Consider a system of $N$ regular and $M$ fractional D3-branes at a toric CY singularity with metric, 
\beqa
ds_{\IY_6}^2 \,=\, dr^2 \, +\, r^2 ds_{\IX_5}^2
\eeqa
The near horizon geometry is a solution of the kind considered in \cite{Klebanov:2000nc} for the conifold and generalized in \cite{Grana:2000jj,Franco:2004jz}, as a particular class of the supersymmetric warped compactification ansatz in \cite{Dasgupta:1999ss,Giddings:2001yu},
\beqa
ds^2 \,=\, Z(r)^{-1/2}\, \eta_{\mu\nu}\, dx^\mu \,dx^\nu\, +\, Z(r)^{1/2} \, [\,  dr^2 \, +\, r^2 ds_{\IX_5}^2\, ].
\eeqa
One obtains a warped version of the singular manifold, which can be regarded as the 5d horizon $\IX_5$ fibered over the 5d space given by 4d Minkowski space and the radial direction $r$. 

There are $M$ units of RR 3-form flux along a non-trivial 3-cycle $\Sigma_3$ (topologically an $\IS^3$ or a Lens space) in $\IX_5$, and a corresponding NSNS 3-form flux, such that the combination (setting the 10d RR axion to zero for simplicity) $G_3=F_3 - \frac i g_s\, H_3$ is a harmonic (2,1)-form, so that the flux is supersymmetric. This $H_3$ flux can be described as a variation in $r$ of the 5d scalar arising from the axion $\phi$ given by the component of the NSNS 2-form $B_2$ along the harmonic 2-form $\omega_2$ Poincar\'{e} dual to $\Sigma_3$ (equivalently, the period of $B_2$ over the 2-cycle $\Sigma_2$ dual to $\Sigma_3$ in $\IX_5$), specifically. 
\beqa
H_3\, =\,  g_s\, M \, \frac {dr}r\, \wedge \omega_2(\Sigma_3)
\label{scalar-variation}
\eeqa
The combination of fluxes is a source of the RR 5-form $dF_5=F_3\wedge H_3$, such that its flux $N$ over $\IX_5$ varies  logarithmically as
\beqa
N \sim  g_s\,  M^2  \ln (r/r_0)\,
\label{n-variation}
\eeqa
where $r_0$ is a cutoff distance. The fluxes also backreact on the geometry, via the warp factor, which obeys
\beqa
(\nabla{^2}_{\IY} Z)\, {\rm vol} (\IY_6)\, =\, g_s\, F_3\wedge H_3
\eeqa
leading to
\beqa
Z(r)\,= \, \, \frac{4 \pi g{_s}^2}{r^4} \, M^2 \, \left( \, \ln \left(\frac{r}{r_0}\right) \, + 1 \,\right) 
\label{warp-variation}
\eeqa
The whole of $\IX_5$ shrinks at $r=0$, but the $F_5$ flux has disappeared by then, so there is no topological obstruction to the shrinking from this side. However,  the 3-cycle $\Sigma_3$ in $\IX_5$ also collapses, and it supports the $F_3$ flux, which is constant. This leads to a naked singularity at the tip of the throat.

The 5d part of the above solution describes what we refer to as a {\em local AdS solution}. It corresponds to a background which locally in $r$ is an AdS$_5$ background, but whose AdS curvature changes in $r$, as in (\ref{warp-variation}). This variation is controlled by that of a 5d scalar, which in the earlier flux throat is $\phi=\int_{\Sigma_2} B_2$, changing from (\ref{scalar-variation}). In purely 5d terms, the defining property for this scalar is that (from the 10d topological coupling $F_3\wedge B_2 \wedge F_5$) it has a 5d topological coupling 
\beqa
S_{\rm CS} \, =\, M \,\phi\, F_5
\eeqa
This is the 5d version of the topological couplings \cite{Dvali:2005an,Kaloper:2008fb}, arising in flux compactifications as described in \cite{Marchesano:2014mla,McAllister:2014mpa}. Upon integrating out the non-dynamical $F_5$, the resulting potential for $\phi$ controls the local (in $r$) value of the vacuum energy. The background value for this 5d field, following from (\ref{scalar-variation}) is
\beqa
d\phi\, =\,  g_s\, M \, \frac {dr}r
\label{variantion-5d}
\eeqa
Alternatively, its boundary condition is fixed by the asymptotic behavior
\beqa
\phi \sim M\, \ln (r/r_0)
\eeqa
The local AdS solution can thus be described as a (in this case, 5d) AdS solution modified by the backreaction of a (5d) scalar $\phi$ with topological coupling to a non-dynamical field strength top-form and obeying (\ref{variantion-5d}). The coupling to the top-form can be replaced by equivalent dual formulations, e.g. the explicit $r$-dependence of the 5d vacuum energy.

The local AdS backgrounds we have described contain a naked singularity at the origin, which in fact is known to admit a smooth deformation (preserving supersymmetry) in certain singularities, starting from the celebrated conifold example \cite{Klebanov:2000hb} and generalized in \cite{Franco:2005fd}. Thus, the local AdS solution should be regarded as defining the asymptotics of certain very general class of warped throats, in principle with or without supersymmetry, and imposing swampland constraints on the possible existence of such throats in quantum gravity. This brings us to the precise formulation of a new swampland conjecture.

\medskip

{\underline {Local AdS-WGC swampland criterion}}:\\
{\em In consistent theory of quantum gravity, there are no stable non-supersymmetric solutions with asymptotics given by local AdS backgrounds, as defined above}.

\subsection{Evidence from deformation and DSB fractional brane systems}
\label{sec:deformation-dsb}

Besides the direct derivation in the spirit of the AdS-WGC, we now present additional support for the local AdS-WGC. Although the following results are known in the literature, their re-interpretation in terms of a swampland constraint is new and provides an interesting insight into the structure of the underlying warped throats and supersymmetry breaking, which we further exploit in later sections.

As mentioned in section \ref{sec:dimer-intro}, there is a large class of local AdS backgrounds arising as holographic duals of (the UV regime of) systems of regular and fractional D3-branes at singularities, specifically, fractional branes of the deformation or DSB kinds ($\NN=2$ fractional branes are discussed in section \ref{sec:ntwo}). We discuss their interplay with the local AdS-WGC in turn.

Toric CY singularities admitting a complex deformation can support deformation branes. The gauge theory on their worldvolumes has an UV RG flow whose holographic dual is given by a supersymmetric local AdS background supported by $M$ units of RR flux on the 3-cycle $\Sigma_3$ associated to the complex deformation. Thus the naked singularity at the origin in the local AdS background can be smoothed out by giving this 3-cycle a finite size. The resulting configuration is a smooth supergravity solution described by a warped version of the deformed CY threefold, preserving supersymmetry, and with asymptotics given by a local AdS background; this is thus in agreement with the local AdS-WGC statement. The field theory counterpart of this deformation process was described in \cite{Klebanov:2000hb,Franco:2005fd}.

Toric CY singularities can also support DSB fractional branes which are not associated to complex deformations. Still, the gauge theory on their worldvolume has a UV RG flow whose holographic dual is a supersymmetric local AdS background supported by $M$ units of RR flux on a 3-cycle $\Sigma_3$. The latter, however, cannot be given a finite size while preserving supersymmetry. Naively, one may think that the infrared region is smoothed out to an alternative configuration breaking supersymmetry, either in the form of a supergravity background beyond the warped CY ansatz (in the spirit of e.g. \cite{Butti:2004pk} in the supersymmetric case), or perhaps involving stringy ingredients, such us explicit sources from branes  or other singular objects. However, if  such re-stabilization would indeed be possible, it would contradict our local AdS-WGC statemetnt. 

The actual answer is that the warped throats created by DSB fractional branes actually do not admit any such stable non-supersymmetric smooth version, in agreement with the local AdS-WGC conjecture. This has actually been already studied in the lilerature, from the gauge theory side. The complex cone over dP$_1$ is the prototypical case of a duality cascade triggered by a DSB brane, and the lack of a supersymmetric vacuum in this dP$_1$ theory was discussed in \cite{Berenstein:2005xa,Franco:2005zu,Bertolini:2005di}. This however does not imply the existence of a non-supersymmetric stable vacuum, rather \cite{Franco:2005zu} already established that the theory shows a runaway behaviour, as follows. By keeping the $U(1)$ factors in the description of the gauge theory, the system has a supersymmetry breaking minimum only if the Fayet-Iliopoulos terms are kept fixed, due to the constraints from the D-term potential. However, the FI terms are actually field dependent, and are controlled by the vevs of closed string twisted sectors. When they are taken as dynamical, the D-term potential can relax in new directions leading to the runaway. The same physics was reinterpreted in \cite{Intriligator:2005aw} as a baryonic runaway direction in the gauge theory with the (massive) $U(1)$'s integrated out. In either of these descriptions, the runaway direction corresponds to a dynamical blow-up of the singularity, since FI terms, or baryonic vevs, are related to blow-up modes. The fractional brane remains as a D5-brane wrapped on a 2-cycle in the dP$_1$ exceptional divisor.

The gravity dual of this runaway has not been determined in the literature, but its structure should correspond to a time-dependent solution, in which the geometry is resolved by growing a finite size dP$_1$ itself, with $M$ explicit D5-branes wrapped on one of its 2-cycles. The latter plays the role of sourcing the $M$ units of RR 3-form, peeling it off the 3-cycle and allowing it to shrink to zero size at the bottom of the (disappearing) throat. 

It is interesting to point out that this system provides an interesting link between two seemingly unconnected swampland criteria. On one hand, the statement that in theories of quantum gravity all FI terms should be field-dependent, and thus dynamical \cite{Komargodski:2009pc}; on the other hand, our newly proposed local AdS-WGC. We expect other connections of the local AdS-WGC constraint with other swampland criteria.

We thus see that the class of throats obtained from the different kinds of fractional branes provide illustrative examples of the local AdS-WGC constraint. In later sections we illustrate the power of this conjecture to exclude candidates to non-supersymmetric throats proposed in the literature.

\subsection{Meta-stable throats}
\label{sec:metastable}

It is important to emphasize that  the present form of the local AdS-WGC still allows for certain forms of non-supersymmetric warped throats. For instance,

$\bullet$ The conjecture poses no conflict so far with the existence of supersymmetry breaking meta-stable throats with local AdS asymptotics. For instance the systems of anti-D3-branes at the bottom of conifold-like  warped throats (i.e. created by deformation fractional branes), extensively used since \cite{Kachru:2003aw}, are in principle allowed \footnote{For discussions on asymptotics and stability of these throats, there is a long-standing debate, see e.g. \cite{Bena:2018fqc} for a recent work, and references therein.}. See also \cite{Kachru:2009kg}, where non-supersymmetric orbifolds are considered and shown to be unstable through nucleation of bubbles of nothing. In contrast with the AdS-WGC, in local AdS throats there is no isometry in the radial direction introducing an infinite volume factor multiplying the decay probability, rather instabilities tend to nucleate near the tip of the throat. Hence, a finite and potentially small decay amplitude is in principle feasible, although this point deserves further study\footnote{We thank M. Montero for raising this point.}.

$\bullet$ Similarly for the nilpotent Goldstino scenario realized in terms of a single anti-D3-brane on top of an O3-plane \cite{Kallosh:2015nia}, for which the stability remarks of \cite{Polchinski:2015bea} specially apply. 

$\bullet$ Finally, global compactifications including warped throats may contain ingredients in the CY bulk which modify non-trivially the boundary conditions in the UV region of the throat, thus changing its asymptotics, and allowing it to evade the local AdS-WGC constraint. For instance, this may well be the case if one introduces euclidean D3-brane instantons on 4-cycles intersecting the underlying DSB D3-brane system (thus, stretching in the radial direction of the throat) to stop their runaway, as proposed in \cite{Florea:2006si} (see also \cite{Blumenhagen:2006xt,Ibanez:2007rs} for related tools). Also, if one includes D7-branes introducing new flavours in DSB D-brane systems, to allow for metastable supersymmetry breaking vacua \cite{Franco:2006es,GarciaEtxebarria:2007vh} in the ISS spirit \cite{Intriligator:2006dd}. For a recent discussion of orientifolded throats, see \cite{Argurio:2017upa}.

\medskip

In the following discussions, we consider several large classes of non-supersymmetric warped throats, and reconcile them with  the local AdS-WGC by looking for decay channels. Whether these decay channels render the configurations unstable or just meta-stable is not constrained by the conjecture in its present form, hence we loosely refer to them as instabilities of the configuration, even in cases where they could host meta-stable backgrounds.

\section{Warped throats with Dynamical Supersymmetry Breaking}
\label{sec:dsb-throat}

In the previous discussion, the system of D3-branes breaking supersymmetry had a fairly manifest runaway behaviour. There are however other systems of D3-branes at singularities which trigger genuine dynamical supersymmetry breaking, rather than runaway. In this section we explore the proposal of embedding such systems in warped throats  \cite{Retolaza:2015nvh}, and how they face the local AdS-WGC.

Again, there are systematic tools for the construction of such theories in terms of D3-branes at toric singularities (possibly in the presence of orientifold quotients), producing $\NN=1$ supersymmetric gauge theories with supersymmetry broken only by non-perturbative dynamics. As explained in \cite{Retolaza:2015nvh}, dimer diagram tools moreover allow to realize them as the theories arising in the infrared of duality cascades of systems of further (deformation) fractional D3-branes at singularities. The gravity dual description of these configurations would correspond to a locally AdS supersymmetric warped throat supported by 3-form fluxes on a 3-cycle associated to a complex deformation, and at whose tip we have the supersymmetry breaking D-brane sector.

If stable, such configurations would lead to a supersymmetry breaking warped throat violating the local AdS-WGC. In this section, we provide a detailed analysis of an illustrative example and show that the configurations are actually unstable. Concretely, although the DSB D3-brane system is consistent in isolation, its embedding into a warped throat contains an instability against bubble nucleation of certain D-brane domain walls. The latter are however more involved than just D3-brane domain walls peeling off the 5-form flux, and provide a novel kind of decay for warped throats. The system also relates to warped throats from (orientifolds of) $\NN=2$ fractional branes, which we discuss as well.

\subsection{The DSB D-brane system}
\label{sec:dsb-gauge}

\begin{figure}[!htp]
\begin{center}

\includegraphics[scale=0.6]{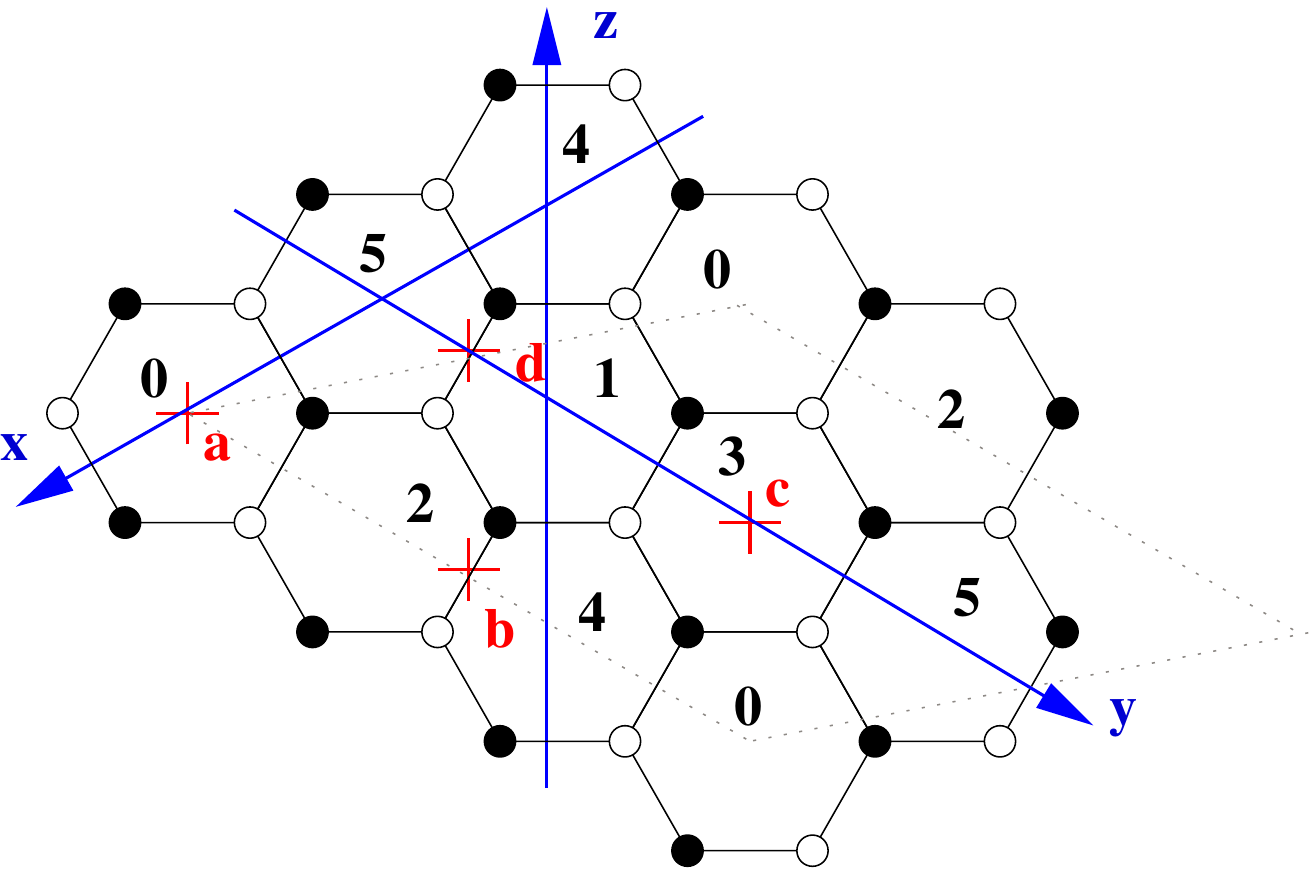}
\caption{Dimer diagram for an orientifold of the $\IC^3/\IZ_6'$ theory, from \cite{Franco:2007ii}.}
\label{dsbZ6}
\end{center}
\end{figure}

To make the discussion concrete, we consider an illustrative explicit example given by the DSB theory introduced in \cite{Franco:2007ii}. We start with the $\IC^3/\IZ_6'$ geometry, where the $\IZ_6'$ generator $\theta$ acts as
\beqa
\theta: \;\; z_i\to e^{2\pi iv_i} z_i
\label{DSB-z6-generator}
\eeqa
with $v=(1,2,-3)/6$. We consider the quotient by an orientifold group $(1+\theta+\ldots +\theta^5) (1+\Omega \alpha (-1)^{F_L})$, where $\alpha$ acts as
\beqa
(z_1,z_2,z_3)\to (e^{2i\pi/12}, e^{4i\pi /12}, e^{-6i\pi /12}).
\eeqa
Equivalently, we may introduce invariant coordinates
\beqa
x=z_1^{\, 6}, ~y=z_2^{\,3}, ~z=z_3^{\, 2}.
\eeqa
in terms of which the orientifold corresponds to the geometric action
\beqa
x\to -x,~ y\to -y, ~z\to -z.
\label{DSB-z6-orientifold-action}
\eeqa
We consider sets of D3-branes at this orientifold singularities. The resulting gauge theory can be determined from its dimer diagram, shown in Figure \ref{dsbZ6}. As discussed in the introduction, there are different choices of orientifold signs, which lead to different results of $SO$ or $Sp$ gauge factors and of $\asymm / \symm$ matter fields. For our choice of interest, corresponding to orientifold signs $(a,b,c,d)=(++--)$, the resulting gauge theory is
\beqa
&SO(n_0)\times U(n_1)\times U(n_2)\times USp(n_3) & \nonumber \\
& (\fund_0,\antifund_1) + (\fund_1,\antifund_2) + (\fund_2,\antifund_3)+& \nonumber \\
& + (\fund_0,\antifund_2) + (\fund_1,\fund_3) + \asymm_2+\antisymm_1 +& \nonumber \\
&+  [\, (\fund_0,\fund_3) + (\fund_1,\fund_2) + (\antifund_1,\antifund_2) \,]&.
\label{DSB-z6-general}
\eeqa

As is familiar \cite{Ibanez:1998qp}, cancellation of non-abelian gauge anomalies is equivalent to the requirement of cancellation of compact RR tadpoles, which leads to
\beqa
-n_0 + n_2+n_3-n_1-4=0.
\eeqa
We consider the solution $n_1=n_3=0$, $n_0=k$, $n_2=k+4$, which yields the gauge group $SO(k)\times U(k+4)$ with 
matter $(\fund,\antifund)+(1,\asymm)$. The $U(1)$ gauge factor is anomalous, with anomaly canceled by Green-Schwarz couplings, which make it massive and  remove it from the massless spectrum. Focusing on  $k=1$, we have an $SU(5)$ theory with chiral multiplets in the $10+{\ov 5}$ and no superpotential. This theory has been argued to show dynamical 
supersymmetry breaking \cite{Affleck:1983vc,Poppitz:1995fh}. Since there is no moduli space, there is an isolated non-supersymmetric vacuum, which however lies at strong coupling and is non-calculable. Nevertheless, the vacuum energy should scale with the strong dynamics scale $\Lambda$ as
\beqa
V\, \sim \, |\Lambda|^4
\label{DSB-energy}
\eeqa
This provides a consistent configuration displaying supersymmetry breaking localized at the tip of the corresponding singularity. 

It is natural to consider its embedding into warped throats, as a possible source of tunable uplifting energy to be used in attempts to build de Sitter string vacua. In the following we argue this not to be possible.

\subsection{The DSB AdS throat}
\label{sec:dsb-ads}

As a warm-up towards such throats, we may consider the simple addition of a large number of dynamical D3-branes to the earlier system, and take the near horizon limit. This corresponds to increasing the rank of all gauge factors in \ref{DSB-z6-general} by the same amount, namely
\beqa
n_0=N+1\quad ,\quad n_1=n_3= N \quad ,\quad n_2=N+ 5
\label{DSB-z6-general-rank}
\eeqa
For consistency with the $USp$ factor, $N$ should be taken even, but is otherwise unconstrained.

Since the DSB D-brane system (including the orientifold and the $k=1$ $SU(5)$ D-brane set) is subleading in $1/N$, standard arguments show that in the large $N$ limit we obtain a gravity dual given by AdS$_5\times \IX$, where $\IX$ corresponds to an orientifold of the $\IZ_6'$ orbifold of $\IS^5$. Note that since the $\IZ_6'$ orbifold contains fixed complex planes in $\IC^3$, there are fixed circles in the action on $\IS^5$. This leads to circles of $\IC^2/\IZ_2$ and $\C^2/\IZ_3$ singularities, which are however well understood \cite{Hanany:1998it,Gukov:1998kk}. The orientifold action (\ref{DSB-z6-orientifold-action}) has instead the origin as only fixed point, hence it is freely acting on $\IS^5$.

At leading order in $1/N$, which corresponds to the classical gravity level, we have a supersymmetric AdS configuration, associated to the near horizon limit of a D-brane system saturating the WGC bound, hence satisfying the AdS-WGC. In the exact configuration, however, the DSB D-brane sector breaks supersymmetry, and implies that at the quantum level the gravitational background becomes non-supersymmetric, hence according to the AdS-WGC, the system should exhibit an instability.

Naively, it would seem that the instability corresponds, as suggested in \cite{Ooguri:2016pdq}, to the emission of shells of D3-branes peeling off the 5-form flux background from the AdS solution. This would correspond, in the underlying picture of D-branes at singularities, to the DSB D-brane system repelling dynamical D3-brane off the origin towards generic points in the transverse space. This actually turns out to be incorrect, as can be shown using the field theory description, using standard supersymmetric field theory arguments. Expelling the dynamical D3-branes corresponds to the Higgsing down the gauge theory with the rank assignment (\ref{DSB-z6-general-rank}) to the original $N=0$ $SU(5)$ theory, by giving vevs to suitable mesonic operators. To make the point, it suffices to turn on a vev for the gauge invariant operator involving fields in the first line in (\ref{DSB-z6-general})
\beqa
\langle\; (\fund_0,\antifund_1) \cdot (\fund_1,\antifund_2) \cdot (\fund_2,\antifund_3)\; \rangle \equiv \Phi^3
\eeqa
Here $\Phi$ is the dimension 1 order parameter for this vev. The superpotential involves only triples of fields from the three different lines in (\ref{DSB-z6-general}), hence it is an F-flat direction. As follows from the D-brane picture, there are more general choices, allowing for three independent vevs -- for similar mesonic operators built from fields in the three different lines in (\ref{DSB-z6-general}) --- for each of the dynamical D3-branes. But for our present purposes it suffices to consider only this overall position vev $\Phi$.

From the viewpoint of the infrared $SU(5)$ theory this corresponds to a Higgsing of the UV $SU(N+5)$ theory by the $N$ flavours acquiring vevs involved in $\Phi$. Denoting $\Lambda$, and $\Lambda_{\rm UV}$ the dynamical scales of the $SU(5)$ and $SU(N+5)$ theories, the potential for $\Phi$ would follow from (\ref{DSB-energy}) from the implicit dependence of the IR scale $\Lambda$ on $\Phi$. However, taking the $SU(5)$ theory, with a $10+{\ov 5}$ matter content, and the UV $SU(N+5)$ theory, with matter content $(3N+1)\antifund + 2N\, \fund + \asymm$, the matching relation is just $\Lambda=\Lambda_{\rm UV}$, with no dependence of $\Phi$. This implies that the DSB D-brane systems does not exert forces on dynamical D3-branes, which are thus not repelled from the origin. The non-supersymmetric AdS configuration is not unstable towards the emission of such D3-brane shells peeling off the 5-form flux.

Actually, the contradiction with the AdS-WGC statement is avoided by a novel mechanism, related to a different kind of instability, which we explain as follows. Let us return to the picture of D3-branes at the orientifold of the $\IC^3/\IZ_6'$ singularity, i.e. the rank assignment (\ref{DSB-z6-general-rank}). The $\IZ_6'$ quotient does not actually define an isolated singularity; indeed, the generator (\ref{DSB-z6-generator}) has the origin as only fixed point, but $\theta^3$ leaves invariant the complex plane parametrized by $z_2$, and $\theta^2$ leaves $z_3$ invariant. This implies that there is a complex plane (along $z_2$) of $\IC^2/\IZ_2$ singularities, and a complex plane (along $z_3$) of $\IC^2/\IZ_3$ singularities. In the field theory, there are flat directions corresponding to splitting some of the dynamical D3-branes into fractional D3-branes (of the $\NN=2$ kind, i.e. D5-branes wrapped on the collapsed cycles of the $\IC^2/\IZ_n$) which can slide off the origin along the corresponding complex plane. Once the non-perturbative supersymmetry breaking kicks in, these flat directions can turn into runaway, providing an instability, bringing back agreement with the AdS-WGC.

The existence of this instability can again be analyzed in terms of the field theory, by Higgsing and scale matching. Consider for concreteness the splitting of dynamical D3-branes into fractional D3-branes of the $\IC^2/\IZ_2$ singularity  associated to $\theta^3$, and moving the latter along $z_2$. A similar analysis could be performed using the fractional branes of the $\IC^2/\IZ_3$ curve of singularities. Motion in $z_2$ corresponds to mesonic vevs for fields in the second line in (\ref{DSB-z6-general}). Denoting the fields $(\fund_0,\antifund_2)$ and $\asymm_2$ by $Q_{A}{}^{i}$ and $A_{ij}$, respectively, and $(\fund_1,\fund_3)$, $\antisymm_1$ by $Q'_{j',B'}$, $S^{i'j'}$, respectively, the vevs for the two kinds of fractional branes have the structure
\beqa
v \, =\, \langle\, \epsilon_{AB}  Q_A{}^i Q_B{}^j A_{ij}\, \rangle\quad ;\quad v'\, =\, \langle\, S^{i'j'}  Q_{i'A} Q_{j'B}  \delta^{AB}\, \rangle
\eeqa
For simplicity we have assumed all fractional D3-branes of the same kind to be located at the same position. The fact that the two different fractional branes are related to vevs of fields Higgsing the combinations of gauge factors (0,2) and (1,3), respectively, is manifest in the dimer diagram in Figure \ref{dsbZ6}, where the above combinations correspond to two sets of faces forming two different strips in the $z_2$ mesonic direction.

Let us compute the scale matching.  Considering for instance $v\gg v'$ (eventually shown to be the realistic regime),  the Higgsing pattern is
\beqa
&& SO(N+1) \times SU(N) \times SU(N+5) \times USp(N)\, \stackrel{v}{\longrightarrow} \\
&&  \stackrel{v}{\longrightarrow} SO(1) \times SU(N) \times SU(5)\times USp(N)\, \stackrel{v'}{\longrightarrow} \,SO(1) \times SU(5) \nonumber
\eeqa
where the $SO(1)$ factor is kept for bookkeeping purposes.
In the first step, the $SU(N+5)$ is Higgsed down to $SU(5)$. In the second step, the $SU(5)$ theory maintains the number of colors, but $2N$ flavours become massive. The scale matching between the IR and UV scales $\Lambda$, $\Lambda_{\rm UV}$ is
\beqa
\Lambda^{13}\, = \, \Lambda_{\rm UV}^{13} \, v'{}^{2N} \, v^{-2N}
\label{scale-match}
\eeqa
Replacing in  (\ref{DSB-energy}), the vev $v$ runs away to infinity, while the vev $v'$ is attracted to zero. 

Note that, although the two kinds of fractional branes have similar features in isolated $\IC^2/\IZ_2$, they have a very different behavior in the presence of the orientifold action. This is in fact manifest already in the orientifold projection on the gauge group and matter content.

The resulting configuration is given by a set of D-branes describing the $SO(1) \times SU(N) \times SU(5)\times USp(N)$ gauge theory. The $SU(5)$ gauge factor still has the antisymmetric matter, but it has extra vector-like flavours, and the theory has supersymmetric vacua \cite{Poppitz:1995fh}. This fits nicely with the vacuum energy from (\ref{DSB-energy}), (\ref{scale-match}) going to zero as $v'\to 0$. Note that the final configuration can be described as a quotient (a $\IZ_3$ orbifold of an orientifold of) a set of $N$ $\NN=2$ fractional branes at $\IC^2/\IZ_2$. This configuration has a supersymmetric gravity dual given by a locally AdS throat of the kind studied in \cite{Bertolini:2000dk,Polchinski:2000mx}. These can be regarded as $\NN=2$ versions of the $\NN=1$ Klebanov-Strassler throats, with the singularity at the origin resolved by a stringy phenomenon, the so-called enhan\c{c}on configuration  \cite{Johnson:1999qt}. The fact that the final end point is a supersymmetric local AdS background avoids conflicts with the local AdS-WGC.

In the gravity picture of the initial configuration, the instability of the non-susy AdS corresponds to the nucleation of bubbles defined by suitable fractional D3-branes, namely D5-branes wrapped on a collapsed $\IP_1$ on the $\IS^1$ of $\IC^2/\IZ_2$ singularities, and with spatial topology $\IS^3$ in the non-compact dimensions, expanding outwards with time. In the interior, we are left with a supersymmetric locally AdS throat induced by the $N$ fractional branes stabilized at the origin, and with the singularity at its tip smoothed out presumably by an enhan\c{c}on configuration.
In contrast with other examples in the literature, this is neither a bubble of nothing nor a bubble removing the 5-form flux completely. It thus corresponds to a novel decay channel for non-supersymmetric warped throats.

\medskip

The $\IC^3/\IZ_6'$ orientifold singularity can be embedded in a locally AdS warped throat associated to a complex deformation, as discussed in section \ref{sec:dsb-throat}. In this setup, supersymmetry breaking on the infrared gauge theory would lead to contradiction with our proposed local AdS-WGC. However, our above analysis of the AdS case shows that the locally AdS throat is already unstable due to D5-brane bubble nucleation (on top of other possible decay channels related to the deformation fractional branes). Hence, the conflict with the local AdS-WGC is solved by the decay channel already solving the potential conflict with the AdS-WGC.

\subsection{Non-supersymmetric warped throats for $\NN=2$ fractional branes}\label{sec:ntwo}

In this section we exploit the previous configuration to obtain a non-trivial example of non-supersymmetric warped throat induced by $\NN=2$ fractional branes. The discussion is straightforward and the arguments should be familiar by now.

Consider the previous orientifold singularity, with D-branes corresponding to the rank assignment 
\beqa
n_0=M+1\quad ,\quad n_1=n_3=0\quad ,\quad n_2=M+5
\eeqa
with $M$ even, for consistency of the (hidden) $USp$ factor.
This leads to a gauge theory with group $SO(M+1)\times SU(M+5)$, with matter $(\fund,\antifund)+(1,\asymm)$. In the limit of large $M$, at leading order we have a gravity dual given by a quotient of the supersymmetric $\NN=2$ warped throats in \cite{Bertolini:2000dk,Polchinski:2000mx}. The configuration is of the local AdS kind, hence the local AdS-WGC constraints should apply.

On the other hand, the gauge theory does not have a supersymmetric vacuum. The $SU(N)$ theory with odd $N$, antisymmetric matter, and no extra flavours, breaks supersymmetry, as shown in \cite{Affleck:1983vc,Affleck:1984xz,Poppitz:1995fh}. Actually, this reference argued for an isolated supersymmetry breaking vacuum for the theory with Yukawa couplings, which remove the classical flat direction. In our present example, such superpotential couplings are absent, and the classical flat direction can turn into runaway ones. This is precisely the conclusion from matching of scales, as in the previous section, which we skip. 

This means that, on the gravity side, the classical background has a decay channel given by nucleation of bubbles of fractional branes, exactly as in the previous section. In this case, however, since there are no fractional branes of the supersymmetry preserving kind, the bubbles completely peel off the 5-form flux background of the configuration leading to a complete decay of the local AdS throat.

This example thus provides an explicit example of the application of the local AdS-WGC constraint to non-supersymmetric warped throats induced by $\NN=2$ fractional branes.

\section{Supersymmetry breaking orientifolds in warped throats}
\label{sec:anti-oplanes}

In the previous sections, we have focused on warped throats whose underlying D-brane configuration is supersymmetric in perturbation theory, with supersymmetry breaking arising from non-perturbative strong dynamics effects. It is interesting to check the behavior of warped throats with more dramatic supersymmetry breaking patterns. In this section, we explore a class of warped throats, where supersymmetry breaking is induced by orientifold planes not preserving the supersymmetry preserved by the CY geometry and the 3-form fluxes. In fact, they correspond to the CPT conjugates of the familiar supersymmetric orientifold planes, so we refer to them as anti-orientifold planes. Systems of anti-orientifold planes in the presence of D-branes are identical to systems of anti-D-branes in the presence of orientifold planes, which have been considered in many non-supersymmetric string constructions, pioneered in \cite{Sugimoto:1999tx,Aldazabal:1999jr,Antoniadis:1999xk,Uranga:1999ib,Angelantonj:2000xf,Rabadan:2000ma}.

\subsection{Non-supersymmetric throats from anti-O3-planes}
\label{sec:anti-oplanes-throats}

We focus on anti-O3-planes in the presence of a large number $N$ of D3-branes, possibly at singularities and with extra $M$ fractional branes. In the underlying D-brane construction, they lead to an explictly non-supersymmetric spectrum, which can be easily determined using open string techniques and (non-supersymmetric projections of) dimer diagrams. For $M=0$, the systems of anti-O3-planes with $N$ D3-branes behave as ``supersymmetric'' and conformal in the leading large $N$ approximation, in the sense that the effects of orientifold planes (noticed via crosscaps) are subleading in the large $N$ limit. This implies that the gravity dual description corresponds to AdS backgrounds which behave as supersymmetric in the classical supergravity approximation, but have supersymmetry breaking effects at 1-loop. Similarly, in systems in the presence of $M$ additional deformation branes, we obtain locally AdS warped throats which are supersymmetric in the leading approximation, but break supersymmetry at the 1-loop order. These AdS and locally AdS configurations thus correspond to classically stable backgrounds, which, if stable in the full theory, would violate the AdS-WGC or the local AdS-WGC, respectively. Our purpose is thus to test the stability of these configurations, providing a check of these conjectures at the quantum level.

Concrete examples are easy to build. For instance, \cite{Kallosh:2015nia} provided tools to embed a single (anti-)O3-plane at the bottom of a warped throat with 3-form fluxes, for instance based on the $xy=z^3w^3$ singularity, a $\IZ_3$ orbifold of the conifold. The deformed conifold itself $xy-zw=t^2$ also admits an involution $(x,y,z,w)\to (y,x,-z,-w)$ leading to O3-planes (in fact, two, located at $z=w=0, x=y=\pm i\,t$) \cite{Garcia-Etxebarria:2015lif}. Considering any of these geometries, we may just replace the O3-planes by anti-O3-planes and obtain explicit locally AdS warped throats with supersymmetry broken by anti-orientifold planes.

\subsection{Dynamics of D3-branes and anti-O3-planes}
\label{sec:anti-oplanes-o3}

It is useful to start considering anti-O3-planes in flat space, in the presence of $N$ D3-branes. In the large $N$ limit, the near horizon limit leads to gravity duals of the form AdS$_5\times \IRP_5$, which behave as supersymmetric at leading order and feel the absence of supersymmetry at order $1/N$. The configuration is the CPT symmetric of O3-planes in the presence of anti-D3-branes (denoted by $\ov{\rm D3}$'s), which was studied in \cite{Uranga:1999ib} following the analysis in \cite{Witten:1998xy} for the supersymmetric O3-D3 system. We now revisit the main points, in anti-O3-plane language.

An anti-O3-plane is a fixed plane of the $\IZ_2$ orientifold action on $\IR^6$, preserving the 16 supersymmetries broken by D3-branes. There are four kinds of anti-O3-planes, classified according to the (discretized) values 0, $\frac 12$ for the NSNS and RR 2-form backgrounds on the $\IRP_2$ (twisted) 2-cycles on the $\IRP_5=\IS^5/\IZ_2$ surrounding the origin in $\IR^6$. In short, comparing with \cite{Witten:1998xy}, the tension of an anti-O3-plane equals that of the corresponding O3-plane, while they have opposite RR charge. The tensions and charges, measured in D3-brane units, for the anti-O3-planes are in the following table.

\begin{center}
\begin{tabular}{|l|c|c|c|}
\hline
D-brane description & $(\theta_{NS},\theta_R)$ & Tension & RR charge \\
\hline\hline 
anti-(O3$^-$)      & $(0,0)$ & -1/2 & $+1/2$ \\
\hline
anti-(O3$^-$) + 1 ${\ov{\rm D3}}$  & $(0,1/2)$ & +1/2 & $-1/2$ \\
\hline
anti- O3$^+$    & $(1/2,0)$ & +1/2 & $-1/2$ \\
\hline
anti- ${\widetilde{\rm O3}}^+$ & $(1/2,1/2)$ & + 1/2 & $-1/2$  \\
\hline
\end{tabular}
\end{center}

Just like for O3-planes, the O3$^-$ is a singlet under the type IIB $SL(2,\IZ)$ and the three remaining ones transform into each other under it. 

The stability of the throats built out using these  anti-O3-plane can be heuristically understood by considering the dynamics of D3-branes in the presence of these anti-O3-planes. Namely, we can consider the previous anti-O3-planes with a $N$  D3-branes on top (as counted in the double cover), and study the stability properties of the system. 

The corresponding analysis can in fact be borrowed from \cite{Uranga:1999ib} (in its CPT conjugate version). It is straightforward to obtain the spectrum of the non-supersymmetric gauge theories on D3-branes in the presence of the different anti-O3-planes. The stability properties of the system can be assessed from the open string perspective, by the computation of the Coleman-Weinberg potential. We instead focus on the dynamics in the dual closed string channel, by comparing the interaction between D3-branes and anti-O3-planes due to exchange in the NSNS and RR channels.
We consider the different cases in turn:

$\bullet$  Consider $N=2p$ D3-branes in the presence of the anti-(O3$^-$). They have opposite sign tensions and equal sign RR charges, hence the gravitational and Coulomb interactions are both repulsive. Thus, D3-branes are expelled away from the anti-(O3$^-$) and the configuration is unstable. 

$\bullet$ Take $N=2p$ D3-branes in the presence of the anti-(O3$^-$) + 1 ${\ov{\rm D3}}$. The D3-branes are attracted to the origin, but when they reach below sub-stringy distances, a tachyon arises from open strings between the stuck ${\ov{\rm D3}}$- and the dynamical D3-branes. The result is a configuration of the anti-(O3$^-$) with one stuck D3-brane at the origin, and $(2p-2)$ dynamical D3-branes. The system at the origin has tension +1/2 and charge +3/2, so the Coulomb repulsion overcomes the gravitational attraction and D3-branes are repelled. The result is a (CPT conjugate) of the nilpotent Goldstino configuration \cite{Kallosh:2015nia}.

$\bullet$ Consider $N=2p$ D3-branes in the presence of the anti-(O3$^+$). The gravitational and Coulomb interactions are both attractive, so the D3-branes are driven to the origin. Contrary to the previous case, however, there is no obvious annihilation between the anti-(O3$^+$) and the D3-branes. This would suggest that the non-supersymmetric AdS$_5\times \IRP_5$ gravity dual is stable, in conflict with the AdS-WGC. Happily, as we will discuss later on, a non-perturbative instability will come to the rescue.

$\bullet$ For  $N=2p$ D3-branes in the presence of the anti-(${\widetilde {\rm O3}}^+$) we have a similar situation. The D3-branes are driven to the origin, and no obvious decay channel seems to be available. This perturbatively stable configuration is however again rendered unstable by a non-perturbative process described later on, thus solving the potential conflict with the AdS-WGC and the local AdS-WGC constraints.

\subsection{Instabilities in throats with anti-O3-planes}
\label{sec:anti-oplanes-o3-throats}

The large $N$ limit of the above configurations of D3-branes on top of anti-O3-planes leads to near horizon geometries classically given by AdS$_5\times\IRP_5$, with $N$ units of RR 5-form flux (as counted in the covering space) and the corresponding discrete NSNS and RR 2-form backgrounds on $\IRP_2\subset\IRP_5$. Absence of supersymmetry is only detectable at the 1-loop (i.e. $1/N$ order), namely via string diagrams involving crosscaps and thus noticing the underlying non-supersymmetric orientifold. Thus, the AdS-WGC condition implies such AdS backgrounds should have instabilities.

The same statement applies in more general local AdS warped throats with anti-O3-planes. For any local AdS warped throat admitting a supersymmetric orientifold involution introducing O3-planes, it is possible to consider the non-supersymmetric version obtained by the introduction of any of the different anti-O3-planes. The resulting gravitational background remains the same at the level of classical supergravity, but subleading corrections encode the breaking of supersymmetry. Thus, the local AdS-WGC conditions imply such local AdS backgrounds should be unstable.

We now analyze the instabilities in these AdS backgrounds, and the same conclusions clearly apply to local AdS configurations. The analysis follows the discussion in the previous section.

$\bullet$ In the case of the anti-(O3$^-$) orientifold projection, the repulsion exerted by the anti-O3-plane on D3-branes translates into a decay channel of the corresponding non-supersymmetric AdS$_5\times \IRP_5$ background, by nucleation of D3-brane bubbles, which discharge the $N$ units of RR 5-form flux, much along the lines suggested in \cite{Ooguri:2016pdq}.

$\bullet$ In the case of the anti-(O3$^-$) with an extra anti-D3-brane, the decay channel of the corresponding non-supersymmetric AdS$_5\times \IRP_5$ background is identical to the previous one, since the two configuration simply differ in the value mod 2 of the RR 5-form flux $N$. Notice that the decay does not change the values of the NSNS and RR 2-form backgrounds, since the anti-(O3$^-$) with either the initial stuck anti-D3-brane or the final stuck D3-brane, both have vanishing NSNS background and non-trivial RR 2-form background.

$\bullet$ In the case of the anti-(O3$^+$) projection, the flat space configuration seems stable. However, the S-dual of the anti-(O3$^+$) is given by the configuration of an anti-(O3$^-$) + 1 ${\ov{\rm D3}}$ of the previous paragraph. This suggests that the  anti-(O3$^+$) can turn into an anti-(O3$^-$) via strong coupling processes. Indeed, notice that if one considers an NS5-brane (whose core is inherently non-perturbative) stretching along three of the anti-O3 directions and three directions transverse to it, the NS5-brane splits the anti-O3 in two halves, which actually have opposite signs for the orientifold plane charge, with one extra half anti-D3-brane on top of the anti-(O3$^-$) half to provide a continuous O3-plane charge across the NS5-brane (see \cite{Elitzur:1997hc} for a review including such brane constructions). This allows to nucleate holes in the anti-(O3$^+$), in whose interior the stuck ${\ov{\rm D3}}$ on the anti-(O3$^-$) can annihilate against one of the D3-branes around it, leading to  repulsion of the remaining D3-branes, and thus, to instability. This suggests that, in the AdS$_5\times\IRP_5$ gravity dual language, there is a decay channel via the nucleation of bubbles bounded by a domain wall given by an NS5-brane wrapped on a maximal $\IRP_2$. From the analysis of topological constraints on wrapped branes in \cite{Witten:1998xy} (derived in the supersymmetric setup, but valid in general), this is indeed allowed. The NS5-brane may moreover carry arbitrarily large D3-brane charge, thus discharging dynamically the RR 5-form flux and rendering the AdS unstable.

$\bullet$ Similar conclusions hold in the case of the anti-($\widetilde{\rm O3}^+$) projection, where now the required domain wall involves a bound state of one NS5- and one D5-brane (aka a (1,1)-fivebrane) wrapped on $\IRP_2\subset \IRP_5$, thus changing both the NSNS and RR 2-form backgrounds. The fivebrane can carry D3-brane charge, so it can peel off the RR 5-form flux of the AdS compactification triggering its instability.

The instabilities of the above non-supersymmetric orientifolds of AdS backgrounds generalize straightforwardly to non-supersymmetric orientifolds of local AdS warped throats. Hence, in this class of examples, the local AdS-WGC is closely related to the ordinary AdS-WGC constraint.

\section{Discussion}
\label{sec:conclu}

In this paper we have proposed a new swampland conjecture forbidding stable non-supersymmetric locally AdS warped throats. This {\em local AdS-WGC} statement generalizes the analogous statement for stable non-supersymmetric AdS vacua. We have illustrated its application, which allows to reinterpret several known results about warped throats from fractional branes, and to derive new results on the (in)stability of large classes of non-supersymmetric throats, with supersymmetry breaking triggered by strong dynamics in infrared D-brane sectors, or by the presence of stringy sources like anti-O3-planes.

Although the local AdS-WGC forbids stable non-supersymmetric throats, it has no direct bearing on meta-stable non-supersymmetric throats. In contrast with the AdS-WGC, there is no isometry in the radial direction introducing an infinite volume factor multiplying the decay probability, so a finite and potentially small decay amplitude is in principle feasible.
The question of whether swampland criteria can impose further restrictions on the meta-stable throats used in dS uplifts is a very interesting one, to which we plan to return in the future.

Several of the instabilities of the non-supersymmetric throats we have discussed are of the runaway kind. In actual 4d compactifications, this corresponds to shortening the throat, thus moderating the hierarchies between the bulk and the throat. Hence, even if the dynamics of the global compactification eventually stabilizes the runaway and renders such configurations more stable, there may remain a question on the tunability of scale hierarchies in the final states. The  possibility that swampland criteria directly constrain such hierarchies is a tantalizing direction we hope to explore in the future.

We have made some interesting progress, and provided yet another hint that the body of knowledge on swampland criteria on effective theories is paving the way towards an era of Quantum Gravitational String Phenomenology.

\section*{Acknowledgments}
We are pleased to thank S. Franco, M. Montero and L. Ib\'anez  for useful discussions. E.G. would like to thank The City College at CUNY for its hospitality during part of this work and S. Franco in particular. This work is partially supported by the grants  FPA2015-65480-P from the MINECO/FEDER, the ERC Advanced Grant SPLE under contract ERC-2012-ADG-20120216-320421 and the grant SEV-2016-0597 of the ``Centro de Excelencia Severo Ochoa" Programme. 

\newpage

\bibliographystyle{JHEP}
\bibliography{mybib}

\providecommand{\href}[2]{#2}\begingroup\raggedright\begin{thebibliography}{10}

\bibitem{delaFuente:2014aca}
A.~de~la Fuente, P.~Saraswat, and R.~Sundrum, {\it {Natural Inflation and
  Quantum Gravity}},  {\em Phys. Rev. Lett.} {\bf 114} (2015), no.~15 151303,
  [\href{http://arxiv.org/abs/1412.3457}{{\tt arXiv:1412.3457}}].

\bibitem{Rudelius:2015xta}
T.~Rudelius, {\it {Constraints on Axion Inflation from the Weak Gravity
  Conjecture}},  {\em JCAP} {\bf 1509} (2015), no.~09 020,
  [\href{http://arxiv.org/abs/1503.00795}{{\tt arXiv:1503.00795}}].

\bibitem{Montero:2015ofa}
M.~Montero, A.~M. Uranga, and I.~Valenzuela, {\it {Transplanckian axions!?}},
  {\em JHEP} {\bf 08} (2015) 032, [\href{http://arxiv.org/abs/1503.03886}{{\tt
  arXiv:1503.03886}}].

\bibitem{Vafa:2005ui}
C.~Vafa, {\it {The String landscape and the swampland}},
  \href{http://arxiv.org/abs/hep-th/0509212}{{\tt hep-th/0509212}}.

\bibitem{Ooguri:2006in}
H.~Ooguri and C.~Vafa, {\it {On the Geometry of the String Landscape and the
  Swampland}},  {\em Nucl. Phys.} {\bf B766} (2007) 21--33,
  [\href{http://arxiv.org/abs/hep-th/0605264}{{\tt hep-th/0605264}}].

\bibitem{ArkaniHamed:2006dz}
N.~Arkani-Hamed, L.~Motl, A.~Nicolis, and C.~Vafa, {\it {The String landscape,
  black holes and gravity as the weakest force}},  {\em JHEP} {\bf 06} (2007)
  060, [\href{http://arxiv.org/abs/hep-th/0601001}{{\tt hep-th/0601001}}].

\bibitem{Ooguri:2016pdq}
H.~Ooguri and C.~Vafa, {\it {Non-supersymmetric AdS and the Swampland}},  {\em
  Adv. Theor. Math. Phys.} {\bf 21} (2017) 1787--1801,
  [\href{http://arxiv.org/abs/1610.01533}{{\tt arXiv:1610.01533}}].

\bibitem{Freivogel:2016qwc}
B.~Freivogel and M.~Kleban, {\it {Vacua Morghulis}},
  \href{http://arxiv.org/abs/1610.04564}{{\tt arXiv:1610.04564}}.

\bibitem{Obied:2018sgi}
G.~Obied, H.~Ooguri, L.~Spodyneiko, and C.~Vafa, {\it {De Sitter Space and the
  Swampland}},  \href{http://arxiv.org/abs/1806.08362}{{\tt arXiv:1806.08362}}.

\bibitem{Harlow:2018jwu}
D.~Harlow and H.~Ooguri, {\it {Constraints on symmetry from holography}},
  \href{http://arxiv.org/abs/1810.05337}{{\tt arXiv:1810.05337}}.

\bibitem{Harlow:2018tng}
D.~Harlow and H.~Ooguri, {\it {Symmetries in quantum field theory and quantum
  gravity}},  \href{http://arxiv.org/abs/1810.05338}{{\tt arXiv:1810.05338}}.

\bibitem{Brennan:2017rbf}
T.~D. Brennan, F.~Carta, and C.~Vafa, {\it {The String Landscape, the
  Swampland, and the Missing Corner}},  {\em PoS} {\bf TASI2017} (2017) 015,
  [\href{http://arxiv.org/abs/1711.00864}{{\tt arXiv:1711.00864}}].

\bibitem{Cheung:2014vva}
C.~Cheung and G.~N. Remmen, {\it {Naturalness and the Weak Gravity
  Conjecture}},  {\em Phys. Rev. Lett.} {\bf 113} (2014) 051601,
  [\href{http://arxiv.org/abs/1402.2287}{{\tt arXiv:1402.2287}}].

\bibitem{Brown:2015iha}
J.~Brown, W.~Cottrell, G.~Shiu, and P.~Soler, {\it {Fencing in the Swampland:
  Quantum Gravity Constraints on Large Field Inflation}},  {\em JHEP} {\bf 10}
  (2015) 023, [\href{http://arxiv.org/abs/1503.04783}{{\tt arXiv:1503.04783}}].

\bibitem{Brown:2015lia}
J.~Brown, W.~Cottrell, G.~Shiu, and P.~Soler, {\it {On Axionic Field Ranges,
  Loopholes and the Weak Gravity Conjecture}},
  \href{http://arxiv.org/abs/1504.00659}{{\tt arXiv:1504.00659}}.

\bibitem{Heidenreich:2015wga}
B.~Heidenreich, M.~Reece, and T.~Rudelius, {\it {Weak Gravity Strongly
  Constrains Large-Field Axion Inflation}},
  \href{http://arxiv.org/abs/1506.03447}{{\tt arXiv:1506.03447}}.

\bibitem{Hebecker:2015rya}
A.~Hebecker, P.~Mangat, F.~Rompineve, and L.~T. Witkowski, {\it {Winding out of
  the Swamp: Evading the Weak Gravity Conjecture with F-term Winding
  Inflation?}},  {\em Phys. Lett.} {\bf B748} (2015) 455--462,
  [\href{http://arxiv.org/abs/1503.07912}{{\tt arXiv:1503.07912}}].

\bibitem{Bachlechner:2015qja}
T.~C. Bachlechner, C.~Long, and L.~McAllister, {\it {Planckian Axions and the
  Weak Gravity Conjecture}},  \href{http://arxiv.org/abs/1503.07853}{{\tt
  arXiv:1503.07853}}.

\bibitem{Junghans:2015hba}
D.~Junghans, {\it {Large-Field Inflation with Multiple Axions and the Weak
  Gravity Conjecture}},  \href{http://arxiv.org/abs/1504.03566}{{\tt
  arXiv:1504.03566}}.

\bibitem{Ibanez:2015fcv}
L.~E. Ibanez, M.~Montero, A.~Uranga, and I.~Valenzuela, {\it {Relaxion
  Monodromy and the Weak Gravity Conjecture}},  {\em JHEP} {\bf 04} (2016) 020,
  [\href{http://arxiv.org/abs/1512.00025}{{\tt arXiv:1512.00025}}].

\bibitem{Hebecker:2015zss}
A.~Hebecker, F.~Rompineve, and A.~Westphal, {\it {Axion Monodromy and the Weak
  Gravity Conjecture}},  {\em JHEP} {\bf 04} (2016) 157,
  [\href{http://arxiv.org/abs/1512.03768}{{\tt arXiv:1512.03768}}].

\bibitem{Heidenreich:2016aqi}
B.~Heidenreich, M.~Reece, and T.~Rudelius, {\it {Evidence for a sublattice weak
  gravity conjecture}},  {\em JHEP} {\bf 08} (2017) 025,
  [\href{http://arxiv.org/abs/1606.08437}{{\tt arXiv:1606.08437}}].

\bibitem{Montero:2016tif}
M.~Montero, G.~Shiu, and P.~Soler, {\it {The Weak Gravity Conjecture in three
  dimensions}},  {\em JHEP} {\bf 10} (2016) 159,
  [\href{http://arxiv.org/abs/1606.08438}{{\tt arXiv:1606.08438}}].

\bibitem{Ibanez:2017oqr}
L.~E. Ibanez, V.~Martin-Lozano, and I.~Valenzuela, {\it {Constraining the EW
  Hierarchy from the Weak Gravity Conjecture}},
  \href{http://arxiv.org/abs/1707.05811}{{\tt arXiv:1707.05811}}.

\bibitem{Gonzalo:2018tpb}
E.~Gonzalo, A.~Herraez, and L.~E. Ibanez, {\it {AdS-phobia, the WGC, the
  Standard Model and Supersymmetry}},  {\em JHEP} {\bf 06} (2018) 051,
  [\href{http://arxiv.org/abs/1803.08455}{{\tt arXiv:1803.08455}}].

\bibitem{Gonzalo:2018dxi}
E.~Gonzalo and L.~E. Ibanez, {\it {The Fundamental Need for a SM Higgs and the
  Weak Gravity Conjecture}},  \href{http://arxiv.org/abs/1806.09647}{{\tt
  arXiv:1806.09647}}.

\bibitem{Ooguri:2017njy}
H.~Ooguri and L.~Spodyneiko, {\it {New Kaluza-Klein instantons and the decay of
  AdS vacua}},  {\em Phys. Rev.} {\bf D96} (2017), no.~2 026016,
  [\href{http://arxiv.org/abs/1703.03105}{{\tt arXiv:1703.03105}}].

\bibitem{Garg:2018reu}
S.~K. Garg and C.~Krishnan, {\it {Bounds on Slow Roll and the de Sitter
  Swampland}},  \href{http://arxiv.org/abs/1807.05193}{{\tt arXiv:1807.05193}}.

\bibitem{Ooguri:2018wrx}
H.~Ooguri, E.~Palti, G.~Shiu, and C.~Vafa, {\it {Distance and de Sitter
  Conjectures on the Swampland}},  \href{http://arxiv.org/abs/1810.05506}{{\tt
  arXiv:1810.05506}}.

\bibitem{Kachru:2003aw}
S.~Kachru, R.~Kallosh, A.~D. Linde, and S.~P. Trivedi, {\it {De Sitter vacua in
  string theory}},  {\em Phys. Rev.} {\bf D68} (2003) 046005,
  [\href{http://arxiv.org/abs/hep-th/0301240}{{\tt hep-th/0301240}}].

\bibitem{Balasubramanian:2005zx}
V.~Balasubramanian, P.~Berglund, J.~P. Conlon, and F.~Quevedo, {\it
  {Systematics of moduli stabilisation in Calabi-Yau flux compactifications}},
  {\em JHEP} {\bf 03} (2005) 007,
  [\href{http://arxiv.org/abs/hep-th/0502058}{{\tt hep-th/0502058}}].

\bibitem{Cicoli:2018kdo}
M.~Cicoli, S.~De~Alwis, A.~Maharana, F.~Muia, and F.~Quevedo, {\it {De Sitter
  vs Quintessence in String Theory}},
  \href{http://arxiv.org/abs/1808.08967}{{\tt arXiv:1808.08967}}.

\bibitem{Kachru:2018aqn}
S.~Kachru and S.~P. Trivedi, {\it {A comment on effective field theories of
  flux vacua}},  \href{http://arxiv.org/abs/1808.08971}{{\tt
  arXiv:1808.08971}}.

\bibitem{Klebanov:2000hb}
I.~R. Klebanov and M.~J. Strassler, {\it {Supergravity and a confining gauge
  theory: Duality cascades and chi SB resolution of naked singularities}},
  {\em JHEP} {\bf 08} (2000) 052,
  [\href{http://arxiv.org/abs/hep-th/0007191}{{\tt hep-th/0007191}}].

\bibitem{Giddings:2001yu}
S.~B. Giddings, S.~Kachru, and J.~Polchinski, {\it {Hierarchies from fluxes in
  string compactifications}},  {\em Phys. Rev.} {\bf D66} (2002) 106006,
  [\href{http://arxiv.org/abs/hep-th/0105097}{{\tt hep-th/0105097}}].

\bibitem{Burgess:2003ic}
C.~P. Burgess, R.~Kallosh, and F.~Quevedo, {\it {De Sitter string vacua from
  supersymmetric D terms}},  {\em JHEP} {\bf 10} (2003) 056,
  [\href{http://arxiv.org/abs/hep-th/0309187}{{\tt hep-th/0309187}}].

\bibitem{Kallosh:2015nia}
R.~Kallosh, F.~Quevedo, and A.~M. Uranga, {\it {String Theory Realizations of
  the Nilpotent Goldstino}},  {\em JHEP} {\bf 12} (2015) 039,
  [\href{http://arxiv.org/abs/1507.07556}{{\tt arXiv:1507.07556}}].

\bibitem{Retolaza:2015nvh}
A.~Retolaza and A.~Uranga, {\it {De Sitter Uplift with Dynamical Susy
  Breaking}},  {\em JHEP} {\bf 04} (2016) 137,
  [\href{http://arxiv.org/abs/1512.06363}{{\tt arXiv:1512.06363}}].

\bibitem{Franco:2005rj}
S.~Franco, A.~Hanany, K.~D. Kennaway, D.~Vegh, and B.~Wecht, {\it {Brane dimers
  and quiver gauge theories}},  {\em JHEP} {\bf 01} (2006) 096,
  [\href{http://arxiv.org/abs/hep-th/0504110}{{\tt hep-th/0504110}}].

\bibitem{Franco:2005sm}
S.~Franco, A.~Hanany, D.~Martelli, J.~Sparks, D.~Vegh, and B.~Wecht, {\it
  {Gauge theories from toric geometry and brane tilings}},  {\em JHEP} {\bf 01}
  (2006) 128, [\href{http://arxiv.org/abs/hep-th/0505211}{{\tt
  hep-th/0505211}}].

\bibitem{Kennaway:2007tq}
K.~D. Kennaway, {\it {Brane Tilings}},  {\em Int. J. Mod. Phys.} {\bf A22}
  (2007) 2977--3038, [\href{http://arxiv.org/abs/0706.1660}{{\tt
  arXiv:0706.1660}}].

\bibitem{Yamazaki:2008bt}
M.~Yamazaki, {\it {Brane Tilings and Their Applications}},  {\em Fortsch.
  Phys.} {\bf 56} (2008) 555--686, [\href{http://arxiv.org/abs/0803.4474}{{\tt
  arXiv:0803.4474}}].

\bibitem{Klebanov:1998hh}
I.~R. Klebanov and E.~Witten, {\it {Superconformal field theory on three-branes
  at a Calabi-Yau singularity}},  {\em Nucl. Phys.} {\bf B536} (1998) 199--218,
  [\href{http://arxiv.org/abs/hep-th/9807080}{{\tt hep-th/9807080}}].

\bibitem{Ibanez:1998qp}
L.~E. Ibanez, R.~Rabadan, and A.~M. Uranga, {\it {Anomalous U(1)'s in type I
  and type IIB D = 4, N=1 string vacua}},  {\em Nucl. Phys.} {\bf B542} (1999)
  112--138, [\href{http://arxiv.org/abs/hep-th/9808139}{{\tt hep-th/9808139}}].

\bibitem{Douglas:1996sw}
M.~R. Douglas and G.~W. Moore, {\it {D-branes, quivers, and ALE instantons}},
  \href{http://arxiv.org/abs/hep-th/9603167}{{\tt hep-th/9603167}}.

\bibitem{Franco:2005zu}
S.~Franco, A.~Hanany, F.~Saad, and A.~M. Uranga, {\it {Fractional branes and
  dynamical supersymmetry breaking}},  {\em JHEP} {\bf 01} (2006) 011,
  [\href{http://arxiv.org/abs/hep-th/0505040}{{\tt hep-th/0505040}}].

\bibitem{Franco:2005fd}
S.~Franco, A.~Hanany, and A.~M. Uranga, {\it {Multi-flux warped throats and
  cascading gauge theories}},  {\em JHEP} {\bf 09} (2005) 028,
  [\href{http://arxiv.org/abs/hep-th/0502113}{{\tt hep-th/0502113}}].

\bibitem{GarciaEtxebarria:2006aq}
I.~Garcia-Etxebarria, F.~Saad, and A.~M. Uranga, {\it {Quiver gauge theories at
  resolved and deformed singularities using dimers}},  {\em JHEP} {\bf 06}
  (2006) 055, [\href{http://arxiv.org/abs/hep-th/0603108}{{\tt
  hep-th/0603108}}].

\bibitem{Becker:1996gj}
K.~Becker and M.~Becker, {\it {M theory on eight manifolds}},  {\em Nucl.
  Phys.} {\bf B477} (1996) 155--167,
  [\href{http://arxiv.org/abs/hep-th/9605053}{{\tt hep-th/9605053}}].

\bibitem{Dasgupta:1999ss}
K.~Dasgupta, G.~Rajesh, and S.~Sethi, {\it {M theory, orientifolds and G -
  flux}},  {\em JHEP} {\bf 08} (1999) 023,
  [\href{http://arxiv.org/abs/hep-th/9908088}{{\tt hep-th/9908088}}].

\bibitem{Grana:2000jj}
M.~Grana and J.~Polchinski, {\it {Supersymmetric three form flux perturbations
  on AdS(5)}},  {\em Phys. Rev.} {\bf D63} (2001) 026001,
  [\href{http://arxiv.org/abs/hep-th/0009211}{{\tt hep-th/0009211}}].

\bibitem{Franco:2004jz}
S.~Franco, Y.-H. He, C.~Herzog, and J.~Walcher, {\it {Chaotic duality in string
  theory}},  {\em Phys. Rev.} {\bf D70} (2004) 046006,
  [\href{http://arxiv.org/abs/hep-th/0402120}{{\tt hep-th/0402120}}].

\bibitem{Berenstein:2005xa}
D.~Berenstein, C.~P. Herzog, P.~Ouyang, and S.~Pinansky, {\it {Supersymmetry
  breaking from a Calabi-Yau singularity}},  {\em JHEP} {\bf 09} (2005) 084,
  [\href{http://arxiv.org/abs/hep-th/0505029}{{\tt hep-th/0505029}}].

\bibitem{Bertolini:2005di}
M.~Bertolini, F.~Bigazzi, and A.~L. Cotrone, {\it {Supersymmetry breaking at
  the end of a cascade of Seiberg dualities}},  {\em Phys. Rev.} {\bf D72}
  (2005) 061902, [\href{http://arxiv.org/abs/hep-th/0505055}{{\tt
  hep-th/0505055}}].

\bibitem{Franco:2007ii}
S.~Franco, A.~Hanany, D.~Krefl, J.~Park, A.~M. Uranga, and D.~Vegh, {\it
  {Dimers and orientifolds}},  {\em JHEP} {\bf 09} (2007) 075,
  [\href{http://arxiv.org/abs/0707.0298}{{\tt arXiv:0707.0298}}].

\bibitem{Maldacena:1997re}
J.~M. Maldacena, {\it {The Large N limit of superconformal field theories and
  supergravity}},  {\em Int. J. Theor. Phys.} {\bf 38} (1999) 1113--1133,
  [\href{http://arxiv.org/abs/hep-th/9711200}{{\tt hep-th/9711200}}]. [Adv.
  Theor. Math. Phys.2,231(1998)].

\bibitem{Ibanez:2017kvh}
L.~E. Ibanez, V.~Martin-Lozano, and I.~Valenzuela, {\it {Constraining Neutrino
  Masses, the Cosmological Constant and BSM Physics from the Weak Gravity
  Conjecture}},  {\em JHEP} {\bf 11} (2017) 066,
  [\href{http://arxiv.org/abs/1706.05392}{{\tt arXiv:1706.05392}}].

\bibitem{Klebanov:2000nc}
I.~R. Klebanov and A.~A. Tseytlin, {\it {Gravity duals of supersymmetric SU(N)
  x SU(N+M) gauge theories}},  {\em Nucl. Phys.} {\bf B578} (2000) 123--138,
  [\href{http://arxiv.org/abs/hep-th/0002159}{{\tt hep-th/0002159}}].

\bibitem{Dvali:2005an}
G.~Dvali, {\it {Three-form gauging of axion symmetries and gravity}},
  \href{http://arxiv.org/abs/hep-th/0507215}{{\tt hep-th/0507215}}.

\bibitem{Kaloper:2008fb}
N.~Kaloper and L.~Sorbo, {\it {A Natural Framework for Chaotic Inflation}},
  {\em Phys. Rev. Lett.} {\bf 102} (2009) 121301,
  [\href{http://arxiv.org/abs/0811.1989}{{\tt arXiv:0811.1989}}].

\bibitem{Marchesano:2014mla}
F.~Marchesano, G.~Shiu, and A.~M. Uranga, {\it {F-term Axion Monodromy
  Inflation}},  {\em JHEP} {\bf 09} (2014) 184,
  [\href{http://arxiv.org/abs/1404.3040}{{\tt arXiv:1404.3040}}].

\bibitem{McAllister:2014mpa}
L.~McAllister, E.~Silverstein, A.~Westphal, and T.~Wrase, {\it {The Powers of
  Monodromy}},  {\em JHEP} {\bf 09} (2014) 123,
  [\href{http://arxiv.org/abs/1405.3652}{{\tt arXiv:1405.3652}}].

\bibitem{Butti:2004pk}
A.~Butti, M.~Grana, R.~Minasian, M.~Petrini, and A.~Zaffaroni, {\it {The
  Baryonic branch of Klebanov-Strassler solution: A supersymmetric family of
  SU(3) structure backgrounds}},  {\em JHEP} {\bf 03} (2005) 069,
  [\href{http://arxiv.org/abs/hep-th/0412187}{{\tt hep-th/0412187}}].

\bibitem{Intriligator:2005aw}
K.~A. Intriligator and N.~Seiberg, {\it {The Runaway quiver}},  {\em JHEP} {\bf
  02} (2006) 031, [\href{http://arxiv.org/abs/hep-th/0512347}{{\tt
  hep-th/0512347}}].

\bibitem{Komargodski:2009pc}
Z.~Komargodski and N.~Seiberg, {\it {Comments on the Fayet-Iliopoulos Term in
  Field Theory and Supergravity}},  {\em JHEP} {\bf 06} (2009) 007,
  [\href{http://arxiv.org/abs/0904.1159}{{\tt arXiv:0904.1159}}].

\bibitem{Bena:2018fqc}
I.~Bena, E.~Dudas, M.~Grana, and S.~Lust, {\it {Uplifting Runaways}},
  \href{http://arxiv.org/abs/1809.06861}{{\tt arXiv:1809.06861}}.

\bibitem{Kachru:2009kg}
S.~Kachru, D.~Simic, and S.~P. Trivedi, {\it {Stable Non-Supersymmetric Throats
  in String Theory}},  {\em JHEP} {\bf 05} (2010) 067,
  [\href{http://arxiv.org/abs/0905.2970}{{\tt arXiv:0905.2970}}].

\bibitem{Polchinski:2015bea}
J.~Polchinski, {\it {Brane/antibrane dynamics and KKLT stability}},
  \href{http://arxiv.org/abs/1509.05710}{{\tt arXiv:1509.05710}}.

\bibitem{Florea:2006si}
B.~Florea, S.~Kachru, J.~McGreevy, and N.~Saulina, {\it {Stringy Instantons and
  Quiver Gauge Theories}},  {\em JHEP} {\bf 05} (2007) 024,
  [\href{http://arxiv.org/abs/hep-th/0610003}{{\tt hep-th/0610003}}].

\bibitem{Blumenhagen:2006xt}
R.~Blumenhagen, M.~Cvetic, and T.~Weigand, {\it {Spacetime instanton
  corrections in 4D string vacua: The Seesaw mechanism for D-Brane models}},
  {\em Nucl. Phys.} {\bf B771} (2007) 113--142,
  [\href{http://arxiv.org/abs/hep-th/0609191}{{\tt hep-th/0609191}}].

\bibitem{Ibanez:2007rs}
L.~E. Ibanez, A.~N. Schellekens, and A.~M. Uranga, {\it {Instanton Induced
  Neutrino Majorana Masses in CFT Orientifolds with MSSM-like spectra}},  {\em
  JHEP} {\bf 06} (2007) 011, [\href{http://arxiv.org/abs/0704.1079}{{\tt
  arXiv:0704.1079}}].

\bibitem{Franco:2006es}
S.~Franco and A.~M.~. Uranga, {\it {Dynamical SUSY breaking at meta-stable
  minima from D-branes at obstructed geometries}},  {\em JHEP} {\bf 06} (2006)
  031, [\href{http://arxiv.org/abs/hep-th/0604136}{{\tt hep-th/0604136}}].

\bibitem{GarciaEtxebarria:2007vh}
I.~Garcia-Etxebarria, F.~Saad, and A.~M. Uranga, {\it {Supersymmetry breaking
  metastable vacua in runaway quiver gauge theories}},  {\em JHEP} {\bf 05}
  (2007) 047, [\href{http://arxiv.org/abs/0704.0166}{{\tt arXiv:0704.0166}}].

\bibitem{Intriligator:2006dd}
K.~A. Intriligator, N.~Seiberg, and D.~Shih, {\it {Dynamical SUSY breaking in
  meta-stable vacua}},  {\em JHEP} {\bf 04} (2006) 021,
  [\href{http://arxiv.org/abs/hep-th/0602239}{{\tt hep-th/0602239}}].

\bibitem{Argurio:2017upa}
R.~Argurio and M.~Bertolini, {\it {Orientifolds and duality cascades:
  confinement before the wall}},  {\em JHEP} {\bf 02} (2018) 149,
  [\href{http://arxiv.org/abs/1711.08983}{{\tt arXiv:1711.08983}}].

\bibitem{Affleck:1983vc}
I.~Affleck, M.~Dine, and N.~Seiberg, {\it {Dynamical Supersymmetry Breaking in
  Chiral Theories}},  {\em Phys. Lett.} {\bf 137B} (1984) 187.

\bibitem{Poppitz:1995fh}
E.~Poppitz and S.~P. Trivedi, {\it {Some examples of chiral moduli spaces and
  dynamical supersymmetry breaking}},  {\em Phys. Lett.} {\bf B365} (1996)
  125--131, [\href{http://arxiv.org/abs/hep-th/9507169}{{\tt hep-th/9507169}}].

\bibitem{Hanany:1998it}
A.~Hanany and A.~M. Uranga, {\it {Brane boxes and branes on singularities}},
  {\em JHEP} {\bf 05} (1998) 013,
  [\href{http://arxiv.org/abs/hep-th/9805139}{{\tt hep-th/9805139}}].

\bibitem{Gukov:1998kk}
S.~Gukov, {\it {Comments on N=2 AdS orbifolds}},  {\em Phys. Lett.} {\bf B439}
  (1998) 23--28, [\href{http://arxiv.org/abs/hep-th/9806180}{{\tt
  hep-th/9806180}}].

\bibitem{Bertolini:2000dk}
M.~Bertolini, P.~Di~Vecchia, M.~Frau, A.~Lerda, R.~Marotta, and I.~Pesando,
  {\it {Fractional D-branes and their gauge duals}},  {\em JHEP} {\bf 02}
  (2001) 014, [\href{http://arxiv.org/abs/hep-th/0011077}{{\tt
  hep-th/0011077}}].

\bibitem{Polchinski:2000mx}
J.~Polchinski, {\it {N=2 Gauge / gravity duals}},  {\em Int. J. Mod. Phys.}
  {\bf A16} (2001) 707--718, [\href{http://arxiv.org/abs/hep-th/0011193}{{\tt
  hep-th/0011193}}]. [,67(2000)].

\bibitem{Johnson:1999qt}
C.~V. Johnson, A.~W. Peet, and J.~Polchinski, {\it {Gauge theory and the
  excision of repulson singularities}},  {\em Phys. Rev.} {\bf D61} (2000)
  086001, [\href{http://arxiv.org/abs/hep-th/9911161}{{\tt hep-th/9911161}}].

\bibitem{Affleck:1984xz}
I.~Affleck, M.~Dine, and N.~Seiberg, {\it {Dynamical Supersymmetry Breaking in
  Four-Dimensions and Its Phenomenological Implications}},  {\em Nucl. Phys.}
  {\bf B256} (1985) 557--599.

\bibitem{Sugimoto:1999tx}
S.~Sugimoto, {\it {Anomaly cancellations in type I D-9 - anti-D-9 system and
  the USp(32) string theory}},  {\em Prog. Theor. Phys.} {\bf 102} (1999)
  685--699, [\href{http://arxiv.org/abs/hep-th/9905159}{{\tt hep-th/9905159}}].

\bibitem{Aldazabal:1999jr}
G.~Aldazabal and A.~M. Uranga, {\it {Tachyon free nonsupersymmetric type IIB
  orientifolds via Brane - anti-brane systems}},  {\em JHEP} {\bf 10} (1999)
  024, [\href{http://arxiv.org/abs/hep-th/9908072}{{\tt hep-th/9908072}}].

\bibitem{Antoniadis:1999xk}
I.~Antoniadis, E.~Dudas, and A.~Sagnotti, {\it {Brane supersymmetry breaking}},
   {\em Phys. Lett.} {\bf B464} (1999) 38--45,
  [\href{http://arxiv.org/abs/hep-th/9908023}{{\tt hep-th/9908023}}].

\bibitem{Uranga:1999ib}
A.~M. Uranga, {\it {Comments on nonsupersymmetric orientifolds at strong
  coupling}},  {\em JHEP} {\bf 02} (2000) 041,
  [\href{http://arxiv.org/abs/hep-th/9912145}{{\tt hep-th/9912145}}].

\bibitem{Angelantonj:2000xf}
C.~Angelantonj, R.~Blumenhagen, and M.~R. Gaberdiel, {\it {Asymmetric
  orientifolds, brane supersymmetry breaking and nonBPS branes}},  {\em Nucl.
  Phys.} {\bf B589} (2000) 545--576,
  [\href{http://arxiv.org/abs/hep-th/0006033}{{\tt hep-th/0006033}}].

\bibitem{Rabadan:2000ma}
R.~Rabadan and A.~M. Uranga, {\it {Type IIB orientifolds without untwisted
  tadpoles, and nonBPS D-branes}},  {\em JHEP} {\bf 01} (2001) 029,
  [\href{http://arxiv.org/abs/hep-th/0009135}{{\tt hep-th/0009135}}].

\bibitem{Garcia-Etxebarria:2015lif}
I.~Garcia-Etxebarria, F.~Quevedo, and R.~Valandro, {\it {Global String
  Embeddings for the Nilpotent Goldstino}},  {\em JHEP} {\bf 02} (2016) 148,
  [\href{http://arxiv.org/abs/1512.06926}{{\tt arXiv:1512.06926}}].

\bibitem{Witten:1998xy}
E.~Witten, {\it {Baryons and branes in anti-de Sitter space}},  {\em JHEP} {\bf
  07} (1998) 006, [\href{http://arxiv.org/abs/hep-th/9805112}{{\tt
  hep-th/9805112}}].

\bibitem{Elitzur:1997hc}
S.~Elitzur, A.~Giveon, D.~Kutasov, E.~Rabinovici, and A.~Schwimmer, {\it {Brane
  dynamics and N=1 supersymmetric gauge theory}},  {\em Nucl. Phys.} {\bf B505}
  (1997) 202--250, [\href{http://arxiv.org/abs/hep-th/9704104}{{\tt
  hep-th/9704104}}].

\end{thebibliography}\endgroup

\end{document}